\def\Box{\kern1pt\vbox{\hrule height 1.2pt\hbox{\vrule width 1.2pt\hskip 3pt
   \vbox{\vskip 6pt}\hskip 3pt\vrule width 0.6pt}\hrule height 0.6pt}\kern1pt}
\def\gtwid{\mathrel{\raise.3ex\hbox{$>$\kern-.75em\lower1ex\hbox{$\sim$}}}}
\def\ltwid{\mathrel{\raise.3ex\hbox{$<$\kern-.75em\lower1ex\hbox{$\sim$}}}}
\def\Box{\kern1pt\vbox{\hrule height 1.2pt\hbox{\vrule width 1.2pt\hskip 3pt
   \vbox{\vskip 6pt}\hskip 3pt\vrule width 0.6pt}\hrule height 0.6pt}\kern1pt}
\documentstyle[epsfig,12pt]{article}

\begin{document}
\begin{titlepage}
\begin{flushright}
hep-ph/0007166 \\ CRETE-00-11 \\ UFIFT-HEP-00-11
\end{flushright}
\begin{center}
\textbf{Back-Reaction In Lightcone QED}
\end{center}
\begin{center}
T. N. Tomaras$^{\dagger}$ and N. C. Tsamis$^{\ddagger}$
\end{center}
\begin{center}
\textit{Department of Physics, and Institute of Plasma Physics \\ University of
Crete and FO.R.T.H. \\ P.O. Box 2208, 710 03 Heraklion, Crete \\ GREECE}
\end{center}
\begin{center}
R. P. Woodard$^*$
\end{center}
\begin{center}
\textit{Department of Physics \\ University of Florida \\ 
Gainesville, FL 32611 \\ UNITED STATES}
\end{center}
\begin{center}
ABSTRACT
\end{center}
\hspace*{.5cm} We consider the back-reaction of quantum electrodynamics upon an
electric field $E(x_+) = - A'_-(x_+)$ which is parallel to $x^3$ and depends 
only on the lightcone coordinate $x_+ = (x^0 + x^3)/\sqrt{2}$. Novel features 
are that the mode functions have simple expressions for arbitrary $A_-(x_+)$, 
and that one cannot ignore the usual lightcone ambiguity at zero $+$ momentum. 
Each mode of definite canonical momenta $k_+$ experiences pair creation at the 
instant when its kinetic momentum $p_+=k_+ - e A_-(x_+)$ vanishes, at which 
point operators from the surface at $x_- =-\infty$ play a crucial role. Our 
formalism permits a more explicit and complete derivation of the rate of 
particle production than is usually given. We show that the system can be 
understood as the infinite boost limit of the analogous problem of an electric 
field which is homogeneous on surfaces of constant $x^0$.
\begin{flushleft}
\end{flushleft}
\begin{flushleft}
$^{\dagger}$ e-mail: tomaras@physics.uoc.gr \\
$^{\ddagger}$ e-mail: tsamis@physics.uoc.gr \\
$^*$ e-mail: woodard@phys.ufl.edu
\end{flushleft}
\end{titlepage}

\section{Introduction}

Many interesting things happen when quantum field theory is formulated on a
non-trivial gauge field or metric background. One of these is that the 
background can cause virtual particles to move so as to engender currents or 
stresses which act to change it. This is the phenomenon of back-reaction. 

Our own fascination with back-reaction concerns a quantum gravitational process
which occurs on an inflating background. Superluminal expansion rips apart 
virtual pairs of gravitons --- or any other effectively massless particle which
is not conformally invariant. Although the total energy of these pairs grows
exponentially with the co-moving time, the corresponding growth of the 3-volume
results in only a constant energy density. The interesting secular effect comes
at the next order when one considers the gravitational potentials engendered by
the pairs. As each pair recedes these potentials remain behind to add with 
those of newly created pairs, and the accumulated gravitational 
self-interaction grows. Because gravity is attractive this self-interaction 
must act to slow inflation. Because gravity is a weak interaction at typical 
inflationary scales, inflation can proceed for a very long time before the 
slowing becomes significant. Because the process is infrared it can be studied 
by naively quantizing general relativity, without regard to that theory's lack 
of perturbative renormalizability. And explicit perturbative computations 
confirm that the slowing effect eventually becomes non-perturbatively strong, 
both for pure gravity \cite{nct1} and for certain scalar models \cite{abm}.

The potential phenomenological implications of this mechanism are staggering. 
It at once provides a realistic model of inflation {\it and} an explanation for
why the currently observed cosmological constant is so small. If one forbids
unnaturally light scalars the model has only a single free parameter --- the 
dimensionless product of Newton's constant and the bare cosmological constant. 
It can therefore make unique and cosmologically testable predictions in a way 
that scalar-driven inflation, with its arbitrary potential, can never do. This
was exploited recently to make predictions for the tensor-to-scalar ratio and
for the tensor and scalar spectral indices of anisotropies in the cosmic 
microwave background \cite{atw}.

There is nonetheless a widespread dissatisfaction with the model. For one 
thing, its most interesting predictions are not easy to infer because they
come after the slowing effect has become strong and perturbation theory has
broken down. Even in the perturbative regime there are well-motivated 
objections to the use of gauge fixed expectation values in the explicit 
computations which have been done \cite{unruh}. On a more subjective level 
there is the feeling that nothing can be understood about quantum gravity 
without first resolving the ultraviolet problem and that the new physics 
behind this should also resolve the problem of the cosmological constant. 
Finally, conventional particle physicists lack intuition about the locally de 
Sitter background in which the process occurs. For all these reasons it is 
interesting to study the phenomenon of back-reaction in a simpler and more 
conventional setting for which there is no doubt either about what happens 
qualitatively or how it can be computed analytically. One such setting is the 
response of quantum electrodynamics to a homogeneous electric field. 

What happens initially when a prepared state is released in the presence of a
homogeneous electric field is that electron-positron pairs emerge from the 
vacuum to form a current which diminishes the electric field. If the state is 
released on a surface of constant $x^0$ with no initial charge then the 
electric field at later times depends only upon $x^0$. This process was 
considered long before the ultraviolet problem of quantum electrodynamics was 
resolved \cite{Klein,Sauter}. Schwinger invented what we now know as the 
in-out background field effective action to compute the rate of particle 
production per unit volume in the presence of a strictly constant electric 
field \cite{Schwinger}. Since then a variety of articles 
\cite{Brezin}--\cite{Kluger2} and monographs \cite{Greiner,Fradkin} have 
treated the issue of what happens when the effect becomes strong.

We cannot hope to add much to the physical picture which has emerged through 
the efforts of so many fine scientists. Indeed, our motive for studying this 
system is that the physics of what happens is {\it not} in doubt. However, we 
do have a technical contribution to make by working out the closely related 
process in which a source-free state is released on a surface of constant 
$x_+ \equiv (x^0 + x^3)/\sqrt{2}$ in the presence of an electric field which
is parallel to $x^3$. The resulting evolution yields a homogeneous electric 
field which depends upon $x_+$ rather than $x^0$. An interesting feature of 
Dirac theory in {\it any} such background is that the mode functions are 
simple. This fact was noted recently by Srinivasan and Padmanabhan 
\cite{Indians1,Indians2} for the special case of a charged scalar in a constant
electric field, although we do not agree with their WKB solution. 

It should be pointed out that our background is not the plane wave treated by 
Wolkow \cite{Wolkow} and Schwinger \cite{Schwinger}. In that background the 
electric field is perpendicular to $x^3$, there is a perpendicular magnetic 
field of the same magnitude, and the two together obey the free Maxwell 
equations. In our background the electric field is {\it parallel} to $x^3$, 
there is no magnetic field, and the free Maxwell equations are only obeyed when
the field is constant. What we have instead is an explicit form for the 
fermion mode functions for a class of backgrounds which is general enough to
include the actual evolution of the electric field as it changes under the 
impact of quantum electrodynamic back-reaction. By taking the expectation value
of the current operator in this general class of backgrounds we obtain the
source term for the effective field equation obeyed by the actual electric
field. This is precisely what we should like to do for quantum gravity in order
to treat the problem of what happens when the slowing effect becomes 
non-perturbatively strong. Therefore many of the same issues of gauge fixing, 
the use of expectation values, renormalization and the breakdown of 
perturbation theory can be examined in a setting where the answer is not in 
doubt.

This paper contains seven sections of which this introduction is the first. In
Section 2 our lightcone coordinate and gauge conventions are stated and we
work out the dynamics of a classical charged particle moving in our general
background. In Section 3 we give a complete operator solution for free QED in 
the presence of this background, expressed in terms of the field operators on 
the surfaces of $x_+ = 0$ and $x_- = -\infty$. It turns out that pair creation
is a discrete event on the lightcone. Each mode passes from positive to 
negative frequency at a certain value of $x_+$ depending upon the mode. At 
this instant each mode experiences a drop in amplitude with the missing 
amplitude taken up by operators from the surface at $x_- = -\infty$. We use
these results in Section 4 to give an explicit, analytic derivation for the 
rate of particle production per unit volume for our general background. In 
Section 5 we compute the one loop expectation value of the current induced by 
such a background. As expected, the ultraviolet divergence resolves itself 
into a renormalization of local terms in Maxwell's equations. Here, as in 
gravity, pair production and back-reaction are infrared effects which can be 
studied without understanding the ultraviolet provided one subtracts the 
divergences and uses the physical couplings in the effective field equations.
A peculiar feature of our one loop result is that back-reaction becomes 
infinitely strong infinitely fast. This is explained in Section 6 by noting
that our lightcone system is the singular, infinite boost limit of the 
traditional system in which the state is prepared on a surface of constant 
$x^0$ and the electric field depends upon $x^0$ rather than $x_+$. Similar 
correspondence limits have been recognized since the earliest work on 
lightcone quantum field theory \cite{Kogut}. Our conclusions comprise Section 
7.

\section{Classical electrodynamics on the lightcone}

All the analysis of this paper is done with a flat, timelike metric. We define
the lightcone coordinates as follows:
\begin{equation}
x_{\pm} \equiv \frac1{\sqrt{2}} \left(x^0 \pm x^3\right) \; . \label{eq:xpm}
\end{equation}
The other (``transverse'') components of $x^{\mu}$ comprise the 2-vector 
$\widetilde{x}$. The same conventions apply to the momentum vector $p^{\mu}$, 
so one might write
\begin{equation}
x^{\mu} p_{\mu} = x^0 p^0 - x^3 p^3 - \widetilde{x} \cdot \widetilde{p} =
x_+ p_- + x_- p_+ - \widetilde{x} \cdot \widetilde{p} \; .
\end{equation}
Note, however, that (\ref{eq:xpm}) results in derivatives with respect to $x_+$
and $x_-$ having their natural expression in terms of derivatives with lowered
indices,
\begin{equation}
\partial_{\pm} = \frac1{\sqrt{2}} \left(\partial_0 \pm \partial_3\right) \; .
\end{equation}
Since we define $\widetilde{\nabla}$ as the transverse components of 
$\partial_{\mu}$ one can write
\begin{equation}
p^{\mu} \partial_{\mu} = p^0 \partial_0 + p^3 \partial_3 + \widetilde{p} \cdot
\widetilde{\nabla} = p_+ \partial_+ + p_- \partial_- + \widetilde{p} \cdot
\widetilde{\nabla} \; .
\end{equation}

We define the lightcone components of the vector potential $A_{\mu}$ in analogy
with those of the derivative operator $\partial_{\mu}$
\begin{equation}
A_{\pm} \equiv \frac1{\sqrt{2}} \left(A_0 \pm A_3\right) \; .
\end{equation}
Our gauge condition is $A_+ = 0$ and we restrict attention to configurations
for which $A_-$ and $\widetilde{A}$ vanish at $x_+ = 0$. This means that only
$A_-$ is ever nonzero, and it depends only upon $x_+$. The nonzero components
of the field strength tensor are
\begin{equation}
F^{30} = - F^{03} = F_{03} = - F_{30} = -A_-^{\prime}(x_+) \; .
\end{equation}
Since we want the electric field, $\vec{E} = \widehat{z} F^{30}$ to be 
initially directed along the positive $z$-axis, it follows that $A_-^{
\prime}(0) < 0$. When necessary, we will therefore assume that $A_-(x_+)$ is a 
{\it decreasing} function of $x_+$. Since the electron's charge is negative 
($e < 0$) our nominal assumption is that $e A_-(x_+)$ is an {\it increasing}
function of $x_+$.

It is instructive to consider the dynamics of a point particle of mass $m$ and 
charge $e < 0$ which moves under the influence of $A_-(x_+)$. From the 
differential form of the Lorentz force law,
\begin{equation}
dp^{\mu} = e F^{\mu\nu} dx_{\nu} \; ,
\end{equation}
we infer the following relations for the lightcone coordinates and momenta:
\begin{eqnarray}
dp_+ & = & -e A_-^{\prime}(x_+) dx_+ \; , \\
dp_- & = & e A_-^{\prime}(x_+) dx_- \; , \\
d\widetilde{p} & = & 0 \; .
\end{eqnarray}
Since $A_-^{\prime}(x_+) dx_+ = dA_-$, the relation for $p_+$ implies that
\begin{equation}
k_+ \equiv p_+(x_+) + e A_-(x_+) \; ,
\end{equation}
is a conserved quantity. Since $dx_- = (p_-/p_+) dx_+$, the relation for $p_-$
implies that the product $p_-(x_+) \times p_+(x_+)$ is also conserved. This
product cannot involve $A_-(x_+)$, because the latter depends upon $x_+$, so 
the correspondence limit in which $A_-$ vanishes determines the mass shell 
relation,
\begin{equation}
2 p_+(x_+) p_-(x_+) = \widetilde{p} \cdot \widetilde{p} + m^2 \equiv 
\widetilde{\omega}^2 \; .
\end{equation}

In the free quantum field theory which corresponds to the motion of such a 
point particle, the conserved quantity $k_+$ is the Fourier conjugate to the
coordinate $x_-$ of the field which creates charge $-e$ and annihilates charge
$e$. We shall follow the convention of Kluger et al. \cite{Kluger} in 
distinguishing between the constant {\it canonical momentum} $k_+$ and the
$x_+$ dependent {\it kinetic momentum} $p_+(x_+) = k_+ - e A_-(x_+)$. We will
also see that 
\begin{equation}
p_-(x_+) = {\widetilde{\omega}^2/2 \over p_+(x_+)} = {\widetilde{\omega}^2/2
\over k_+ - e A_-(x_+)} \; , \label{eq:p-}
\end{equation}
is indeed the eigenvalue of the operator $i\partial_+$. A fact of crucial
importance is that it changes sign when $p_+(x_+)$ passes through zero.

We conclude by following the trajectory of a point particle of mass $m$ and 
charge $e < 0$ as it moves under the influence of $A_-(x_+)$. Since $dx_- =
(p_-/p_+) dx_+$ we can integrate to find
\begin{equation}
x_-(x_+) = x_-(0) + \int_0^{x_+} {\frac12 \widetilde{\omega}^2 du \over [k_+ -
e A_-(u)]^2} \; .
\end{equation}
Under our nominal assumption that $e A_-(u)$ is an increasing function, $k_+ -
e A_-(u)$ must pass through zero at some value $u_{\rm crit} > 0$, at least for 
modes whose initial momentum $k_+$ is small. The integral diverges if $k_+ - e 
A_-(u)$ goes to zero even as fast as $\sqrt{u_{\rm crit} - u}$ --- and note 
that $e A_-(x_+)$ is growing {\it linearly} at $x_+ = 0$. What this divergence
means physically is that the electron accelerates to the speed of light and 
leaves the manifold moving parallel to the $x_-$ axis as shown on Fig.~1. 

\begin{figure}
\centerline{\epsfig{file=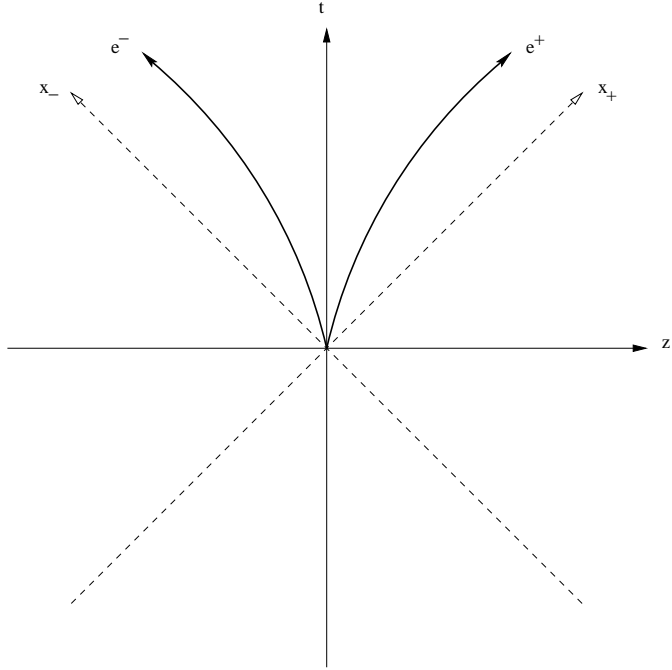,height=3.5in}}
\caption{The evolution of an $e^+ e^-$ pair created at $x_+ = x_- = 0$.}
\end{figure}

The result for positrons is obtained by simply changing $e$ to $-e$. Note that 
although positrons also accelerate to the speed of light they move parallel to 
the $x_+$ axis and do not leave the manifold. We can therefore anticipate that,
for $E(x_+) > 0$, pair creation on the lightcone manifests itself by the 
accumulation of a charge density of positrons whose electron partners have left
the manifold. Since electrons exit the manifold by reaching the speed of light 
we can also anticipate that they induce an infinite current. These suspicions 
will be confirmed by the detailed calculations of Sections 4 and 5. Why the 
lightcone must show an infinite effect will be explained by the correspondence 
limit of Section 6.

\section{QED on the lightcone}

The lightcone components of the gamma matrices are
\begin{equation}
\gamma_{\pm} \equiv \frac1{\sqrt{2}} \left(\gamma^0 \pm \gamma^3\right) \; .
\end{equation}
Note that $(\gamma_{\pm})^2 = 0$. We follow Kogut and Soper \cite{Kogut} in
defining lightcone spinor projection operators,
\begin{equation} 
P_{\pm} \equiv \frac12 \left(I \pm \gamma^0 \gamma^3\right) = \frac12 
\gamma_{\mp} \gamma_{\pm} \; .
\end{equation}
These act on the Dirac bispinor to give its ``$+$'' and ``$-$'' components,
\begin{equation}
\psi_{\pm} \equiv P_{\pm} \psi \qquad , \qquad \psi^{\dagger}_{\pm} \equiv
\psi^{\dagger} P_{\pm} \; .
\end{equation}
It is convenient to Fourier transform on the transverse coordinates,
\begin{equation}
\widetilde{\psi}_{\pm}\left(x_+,x_-,\widetilde{k}\right) \equiv \int d^2
\widetilde{x} e^{-i\widetilde{k} \cdot \widetilde{x}} \psi_{\pm}\left(x_+,x_-,
\widetilde{x}\right) \; .
\end{equation}
Note that the transverse derivative operator $\widetilde{\nabla}$ becomes 
$i \widetilde{k}$ in the Fourier representation. Because transverse
coordinates play so little role we shall often omit $\widetilde{k}$ from
the argument list to simplify the notation.

With these conventions the Dirac equation becomes
\begin{equation}
\left(\gamma^{\mu} i\partial_{\mu} - \gamma^{\mu} e A_{\mu} - m\right) 
\widetilde{\psi} = \left(\gamma_+ i \partial_+ + \gamma_- (i\partial_- - 
e A_-) - \widetilde{\gamma} \cdot \widetilde{k} - m \right) 
\widetilde{\psi} \; ,
\end{equation}
where it should be noted that $e = -|e|$ is the charge of the electron. 
Multiplication alternately with $\gamma_-$ and $\gamma_+$ gives 
\begin{eqnarray}
i\partial_+ \widetilde{\psi}_+(x_+,x_-) & = & \left(m - \widetilde{\gamma} 
\cdot \widetilde{k} \right) \frac12 \gamma_- \widetilde{\psi}_-(x_+,x_-)
\; , \label{eq:psi+} \\
\left({\mbox{} \over \mbox{}} i\partial_- - e A_-(x_+)\right) 
\widetilde{\psi}_-(x_+,x_-) & = & \left(m - \widetilde{\gamma} \cdot 
\widetilde{k} \right) \frac12 \gamma_+ \widetilde{\psi}_+(x_+,x_-) \; . 
\label{eq:psi-}
\end{eqnarray}
One can integrate (\ref{eq:psi+}) from the initial value surface at $x_+ =
0$,
\begin{equation}
\widetilde{\psi}_+(x_+,x_-) = \widetilde{\psi}_+(0,x_-) - \int_0^{x_+} du
\left(m - \widetilde{\gamma} \cdot \widetilde{k}\right) \frac{i}2 \gamma_- 
\widetilde{\psi}_-(u,x_-) \; .
\end{equation}
A similar integration of (\ref{eq:psi-}) from the surface at $x_- = -L$ 
can be achieved by multiplying with $e^{ie A_- x_-}$,
\begin{eqnarray}
\lefteqn{\widetilde{\psi}_-(x_+,x_-) = e^{-i e A_-(x_+) (x_- + L)} 
\widetilde{\psi}_-(x_+,-L)} \nonumber \\
& & - e^{-i e A_-(x_+) x_-} \int_{-L}^{x_-} dv e^{i e A_-(x_+) v}
\left(m - \widetilde{\gamma} \cdot \widetilde{k}\right) \frac{i}2 \gamma_+
\widetilde{\psi}_+(x_+,v) \; .
\end{eqnarray}
Substituting this into the previous equation for $\widetilde{\psi}_+$ and
iterating gives the complete initial value solution for $\widetilde{\psi}_+$
on the region $x_+ > 0$ and $x_- > -L$:
\begin{eqnarray}
\lefteqn{\widetilde{\psi}_+(x_+,x_-) = \sum_{n=0}^{\infty} \left(-\frac12
\widetilde{\omega}^2\right)^n \int_0^{x_+} du_1 e^{-i e A_-(u_1) x_-} 
\int_{-L}^{x_-} dv_1 e^{i e A_-(u_1) v_1} \dots} \nonumber \\
& & \dots \int_0^{u_{n-1}} du_n e^{-i e A_-(u_n) v_{n-1}} \int_{-L}^{v_{n-1}}
dv_n e^{i e A_-(u_n) v_n} \left\{{\mbox{} \over \mbox{}} {\widetilde \psi}_+(0,
v_n) \right. \nonumber \\
& & -\left. \int_0^{u_n} du e^{-i A_-(u) v_n} \left(m- \widetilde{\gamma} \cdot 
\widetilde{k}\right) \frac{i}2 \gamma_- e^{-i e A_-(u) L} \widetilde{\psi}_-(u,
-L)\right\} \; . \label{eq:series}
\end{eqnarray}
A similar expansion for $\widetilde{\psi}_- = \widetilde{\omega}^{-2} 
(m - \widetilde{\gamma} \cdot \widetilde{k}) \gamma_+ i \partial_+ 
\widetilde{\psi}_+$ follows from (\ref{eq:psi+}).

Of course we are interested in the limit as $L$ becomes infinite, in which
case the series (\ref{eq:series}) can be summed. For $n > 0$ we first extend
the integration over $v_n$ to the full real line using the identity:
\begin{equation}
\theta(v_{n-1} - v_n) e^{i e A_-(u_n) [v_n - v_{n-1}]} = \int_{-\infty}^{
\infty} {dk_+ \over 2 \pi} {i e^{i (k_+ + i \epsilon) [v_n - v_{n-1}]} \over 
k_+ - e A_-(u_n) + i \epsilon} \; .
\end{equation}
Owing to the factor of $e^{-\epsilon v_n}$ the integration over $v_n$ only 
makes sense provided the integration over $k_+$ is done first. To change the 
order of integration one must appropriately regulate the lower limit,
\begin{equation}
\int_{-\infty}^{\infty} dv_n F(v_n) \int_{-\infty}^{\infty} {dk_+ \over 2\pi}
G(k_+) = \lim_{\epsilon \rightarrow 0^+}\int_{-\infty}^{\infty} {dk_+ \over 2
\pi} G(k_+) \int_{-1/\epsilon}^{\infty} dv F(v) \; .
\end{equation}
The limit $\epsilon \rightarrow 0^+$ will be understood in all subsequent
expressions, as per the usual convention (for a different $\epsilon$) in 
quantum field theory.

The next step is to move the $k_+$ integration all the way to the left and
perform the integrations over $v_i$ successively, from $i = n-1$ to $i = 1$, 
using,
\begin{equation}
\int_{-\infty}^{v_{i-1}} dv_i e^{-i [k_+ - e A_-(u_i) + i \epsilon] v_i} =
{i e^{-i [k_+ - e A_-(u_i) + i \epsilon] v_{i-1}} \over k_+ - e A_-(u_i)
+ i \epsilon} \; .
\end{equation}
Since the integrand at this stage is the product over the same function of 
each $u_i$ --- $f(u_i) \equiv [k_+ - e A_-(u_i) + i \epsilon]^{-1}$ --- one 
can factor the $u_i$ integrations,
\begin{eqnarray}
\int_0^{x_+} du_1 f(u_1) \dots \int_0^{u_{n-1}} du_n f(u_n) & = & \frac1{n!} 
\left[ \int_0^{x_+} du_1 f(u_1)\right]^n \; , \\
\int_0^{x_+} du_1 f(u_1) \dots \int_0^{u_n} du g(u) & = & \int_0^{x_+} du 
\frac{g(u)}{n!} \left[ \int_u^{x_+} du_1 f(u_1) \right]^n \; . \quad
\end{eqnarray}
The $n=0$ term can be included using the Fourier inversion theorem,
\begin{equation}
h(x_-) = \int_{-\infty}^{\infty} {dk_+ \over 2\pi} e^{-i (k_+ + i \epsilon) 
x_-} \int_{-1/\epsilon}^{\infty} dv e^{i (k_+ + i\epsilon) v} h(v) \; .
\end{equation}
The resulting series gives an exponential. For the terms proportional to 
$\widetilde{\psi}_-$ we get,
\begin{eqnarray}
\lefteqn{\sum_{n=0}^{\infty} \frac1{n!} \left[-\frac{i}2 \widetilde{\omega}^2 
\int_u^{x_+} {du_1 \over k_+ - e A_-(u_1) + i \epsilon}\right]^n} \nonumber \\
& & = \exp\left[- \frac{i}2 \widetilde{\omega}^2 \int_u^{x_+} {du_1 \over k_+ 
- e A_-(u_1) + i \epsilon}\right] \equiv {\cal E}[A_-]\left(u,x_+;k_+,
\widetilde{k} \right) \; . \label{eq:E[A]}
\end{eqnarray}
The terms proportional to $\widetilde{\psi}_+$ give ${\cal E}[A_-]\left(0,x_+;
k_+,\widetilde{k}\right)$. 

It remains to perform the final integration over $v$. For the terms 
proportional to $\widetilde{\psi}_+$ this gives our $\epsilon$-regulated 
Fourier transform,
\begin{equation}
\Xi_0\left(k_+,\widetilde{k}\right) \equiv \int_{-1/\epsilon}^{\infty} dv 
e^{i (k_+ + i \epsilon) v} \widetilde{\psi}_+\left(0,v,\widetilde{k}\right)\; .
\end{equation}
For the terms proportional to $\widetilde{\psi}_-$ the integral over $v$
results in a delta sequence,
\begin{equation}
\Delta\left(k_+ - e A_-(u);\epsilon\right) \equiv {i e^{-i[k_+ - e A_-(u) + i 
\epsilon]/\epsilon} \over k_+ - e A_-(u) + i \epsilon} \; ,
\end{equation}
whose distributional limit would be $2\pi \delta(k_+ - e A_-)$ if it were
multiplied by a test function. The final result is,
\begin{eqnarray}
\lefteqn{\widetilde{\psi}_+\left(x_+,x_-,\widetilde{k}\right)} \nonumber \\
& & = \int_{-\infty}^{\infty} {dk_+ \over 2\pi} e^{-i (k_+ + i \epsilon) x_-} 
\left\{ {\cal E}[A_-]\left(0,x_+;k_+,\widetilde{k}\right) \Xi_0\left(k_+,
\widetilde{k}\right) \right. \nonumber \\
& & \qquad \left. - \int_0^{x_+} du \Delta\left(k_+ - e A_-(u);\epsilon\right) 
{\cal E}[A_-]\left(u,x_+;k_+,\widetilde{k}\right) \Phi_{\infty}\left(u,
\widetilde{k}\right) \right\} \; , \quad \label{eq:solution}
\end{eqnarray}
where we define,
\begin{equation}
\Phi_{\infty}\left(u,\widetilde{k}\right) \equiv \lim_{L \rightarrow \infty} 
\left(m - \widetilde{\gamma} \cdot \widetilde{k}\right) \frac{i}2 \gamma_- 
\widetilde{\psi}_-\left(u,-L,\widetilde{k}\right) e^{-ie A_-(u) L} \; .
\end{equation}
Because the factor of ${\cal E}[A_-]\left(u,x_+;k_+,\widetilde{k}\right)$ 
develops a singular phase as $k_+$ approaches $e A_-(u)$, the distributional
limit of the delta sequence in the second term must be taken with care. We 
shall postpone this to the next section.

It is worth commenting on two exceptional properties of our solution 
(\ref{eq:solution}). First, it is valid {\it for arbitrary} vector potential
$A_-(x_+)$. If the state at $x_+ = 0$ is translation invariant in $x_-$ and
$\widetilde{x}$ then back-reaction will change the way $A_-$ depends upon $x_+$
but it cannot induce other potentials or dependence upon other coordinates. Of
course the photon propagator is not affected by the background, nor are the
vertices. So we can evaluate the expectation value of the current operator --- 
to as high an order in the loop expansion as is desired --- for a class of 
vector potentials which certainly includes the actual solution. The only 
additional simplification one would obtain by making the electric field 
constant ($A_-(x_+) = - E x_+$) is that then the integral over $u_1$ in the 
mode functions (\ref{eq:E[A]}) can be explicitly performed. We shall see, in 
Sections 4 and 5, that this is not required in order to be able to compute 
either the rate of particle production or the expectation value of the current 
operator.

The second property is that our $i \epsilon$ prescription provides a precise
definition for the ambiguity at zero $+$ momentum which, for $m \neq 0$ and/or 
more than two spacetime dimensions, is traditionally left unresolved in 
lightcone quantum field theory. (See, for example, footnote \#12 in the work of
Kogut and Soper \cite{Kogut}.) One can usually avoid doing this because the 
analyticity of scattering amplitudes permits one to infer the zero momentum 
limit from the result for nonzero momentum. In our background the problem is 
aggravated by the fact that {\it every} mode with positive canonical momentum 
$k_+$ becomes singular when its kinetic momentum $p_+(x_+) = k_+ - e A_-(x_+)$ 
passes through zero. At this instant the mode functions ${\cal E}[A_-]\left(0,
x_+;k_+,\widetilde{k}\right)$ oscillate with infinite rapidity and one requires
the $i \epsilon$ prescription to precisely define what happens. Note too that 
we have {\it derived} it rather than simply making an {\it ah hoc} guess. As an 
essential part of the derivation we have found that $\psi_+\left(x_+,x_-,
\widetilde{x}\right)$ is determined not just by $\psi_+\left(0,x_-,\widetilde{
x}\right)$ but also by $\psi_-\left(x_+,-\infty,\widetilde{x}\right)$. When
$A_- = 0$ (and $m \neq 0$ and/or the number of spacetime dimensions is greater
than two) one can ignore the data from the surface at $x_- = -\infty$ because 
it remains segregated in the $k_+ = 0$ mode whose contribution to scattering 
processes is inferred by analytically continuing the result from $k_+ \neq 0$. 
We shall see in the next section that this data cannot be ignored in our 
background and that it plays an essential role in the process of particle 
production.

To complete our operator construction of free Dirac theory in the presence of 
$A_-(x_+)$ we must specify how the fundamental operators $\Xi_0\left(k_+,
\widetilde{k}\right)$ and $\Phi_{\infty}\left(u,\widetilde{k}\right)$ act upon 
one another. Of course the operator algebra derives from canonical 
quantization. The Fourier transform (in $\widetilde{x}$) of the Dirac 
Lagrangian is,\footnote{Note that the quantity $\widetilde{\psi}^{\dagger}$ is 
computed by Fourier transforming {\it first} and then taking the adjoint.}
\begin{eqnarray}
{\cal L} & = & \widetilde{\psi}^{\dagger} \gamma^0 \left(\gamma^{\mu} i
\partial_{\mu} - \gamma^{\mu} e A_{\mu} - m\right) \widetilde{\psi} \; , \\
& = & \sqrt{2} \widetilde{\psi}^{\dagger}_+ \left[i\partial_+ \widetilde{
\psi}_+ - (m - \widetilde{\gamma} \cdot \widetilde{k}) \frac12 \gamma_- 
\widetilde{\psi}_-\right] \nonumber \\
& & \mbox{} + \sqrt{2} \widetilde{\psi}^{\dagger}_- \left[(i\partial_- - e A_-)
\widetilde{\psi}_- - (m - \widetilde{\gamma} \cdot \widetilde{k}) \frac12 
\gamma_+ \widetilde{\psi}_+\right] \; .
\end{eqnarray}
The variable conjugate to $\widetilde{\psi}_+$ under $x_+$ evolution is $i 
\sqrt{2} \widetilde{\psi}^{\dagger}_+$, so we must have,
\begin{equation}
\left\{\widetilde{\psi}_+(x_+,x_-,\widetilde{k}),\widetilde{\psi}^{\dagger}_+(
x_+,y_-,\widetilde{q})\right\} = \frac1{\sqrt{2}} P_+ \delta(x_- - y_-) 
(2 \pi)^2 \delta^2(\widetilde{k} - \widetilde{q}) \; . 
\label{eq:anticom+}
\end{equation}
Since the variable conjugate to $\widetilde{\psi}_-$ under $x_-$ evolution is 
$i \sqrt{2} \widetilde{\psi}^{\dagger}_-$, we must similarly have,
\begin{equation}
\left\{\widetilde{\psi}_-(x_+,x_-,\widetilde{k}),\widetilde{\psi}^{\dagger}_-(
y_+,x_-,\widetilde{q})\right\} = \frac1{\sqrt{2}} P_- \delta(x_+ - y_+) 
(2 \pi)^2 \delta^2(\widetilde{k} - \widetilde{q}) \; . 
\label{eq:anticom-}
\end{equation}
Operators on an arbitrary surface of constant $x_+$ do {\it not} generally
anti-commute with those on an arbitrary surface of constant $x_-$. However, by
causality we know that the operators at $x_+ = 0$ do anti-commute with those
at $x_- = -\infty$. So the only nonzero anti-commutators among the fundamental
operators are:
\begin{eqnarray}
\left\{\Xi_0(k_+,\widetilde{k}),\Xi_0^{\dagger}(q_+,\widetilde{q})\right\} &= &
\frac1{\sqrt{2}} P_+ (2 \pi)^3 \delta(k_+ - q_+) \delta^2(\widetilde{k} - 
\widetilde{q}) \; , \label{eq:Psicom} \\
\left\{\Phi_{\infty}(x_+,\widetilde{k}),\Phi_{\infty}^{\dagger}( y_+,
\widetilde{q})\right\} & = & \frac{\widetilde{\omega}^2}{2 \sqrt{2}} P_+ 
\delta(x_+ - y_+) (2 \pi)^2 \delta^2(\widetilde{k} - \widetilde{q}) \; . 
\label{eq:Phicom}
\end{eqnarray}

\section{Particle production on the lightcone}

Equation (\ref{eq:solution}) expresses the free field $\widetilde{\psi}_+\left(
x_+,x_-,\widetilde{k}\right)$ in terms of the fundamental operators
$\Xi_0\left(k_+,\widetilde{k}\right)$ and $\Phi_{\infty}\left(u,\widetilde{k}
\right)$. We have just seen in (\ref{eq:Psicom}-\ref{eq:Phicom}) how these 
fundamental operators act upon one another and upon their adjoints. Their 
particle interpretation in free field theory derives from the lightcone 
``Hamiltonian'' --- that is, from the generator of $x_+$ evolution. Since the 
Dirac Lagrangian vanishes as a consequence of the field equations the 
Hamiltonian density is just the $p \dot{q}$ term,
\begin{equation}
{\cal H}\left(x_+,x_-,\widetilde{x}\right) = \sqrt{2} \psi_+^{\dagger}\left(
x_+,x_-,\widetilde{x}\right) i \partial_+ \psi_+\left(x_+,x_-,\widetilde{x}
\right) \; .
\end{equation}
The Hamiltonian is the integral of this over $\widetilde{x}$ and our 
$\epsilon$-truncated portion of the $x_-$ axis. We can express it in terms of
$\widetilde{\psi}_+\left(x_+,x_-,\widetilde{k}\right)$ using Parseval's 
theorem,
\begin{equation}
H(x_+) = \int_{-1/\epsilon}^{\infty} dx_- \int {d^2\widetilde{k}\over (2\pi)^2}
\sqrt{2} \widetilde{\psi}_+^{\dagger}\left(x_+,x_-,\widetilde{k}\right) i
\partial_+ \widetilde{\psi}_+\left(x_+,x_-,\widetilde{k}\right) \; .
\end{equation}

As might have been expected from this system's invariance under translations
in $x_-$ and $\widetilde{x}$, the Hamiltonian becomes diagonal in momentum 
space. To see this we take the field's $\epsilon$-regulated Fourier transform 
on $x_-$,
\begin{eqnarray}
\lefteqn{\Psi\left(x_+,k_+,\widetilde{k}\right) \equiv \int_{-1/\epsilon}^{
\infty} dx_- e^{i (k_+ + i \epsilon) x_-} \widetilde{\psi}_+\left(x_+,x_-,
\widetilde{k} \right) \; ,} \\
& & = {\cal E}[A_-]\left(0,x_+;k_+,\widetilde{k}\right) \Xi_0\left(k_+,
\widetilde{k}\right) \nonumber \\
& & \qquad - \int_0^{x_+} du \Delta\left(k_+ - e A_-(u);\epsilon\right) 
{\cal E}[A_-]\left(u,x_+;k_+,\widetilde{k}\right) \Phi_{\infty}\left(u,
\widetilde{k}\right) \; . \label{eq:Psik}
\end{eqnarray}
In the limit of small $\epsilon$ the Hamiltonian becomes,
\begin{equation}
H(x_+) = \int_{-\infty}^{\infty} {dk_+ \over 2\pi} \int {d^2\widetilde{k} \over
(2\pi)^2} \sqrt{2} \Psi^{\dagger}\left(x_+,k_+,\widetilde{k}\right) i\partial_+
\Psi\left(x_+,k_+,\widetilde{k}\right) \; .
\end{equation}

This last expression for $H(x_+)$ implies that the $x_+$-dependent ``energy'' 
carried by $\Psi\left(x_+,k_+,\widetilde{k} \right)$ is its eigenvalue under $-
i \partial_+$. From the first term of (\ref{eq:Psik}) we see that, if 
$\Psi\left(x_+,k_+,\widetilde{k}\right)$ is an eigenfunction of $-i\partial_+$,
its eigenvalue must be,
\begin{equation}
-i \partial_+ \ln\left\{ {\cal E}[A_-]\left(0,x_+;k_+,\widetilde{k}\right)
\right\} = {-\widetilde{\omega}^2/2 \over k_+ - e A_-(x_+) + i \epsilon} \; .
\end{equation}
When $\epsilon$ vanishes this is precisely minus the result (\ref{eq:p-}) we 
found at the end of Section 2 for the $p_-$ momentum of a classical charged 
particle moving in our vector potential. We therefore expect $\Psi\left(x_+,
k_+,\widetilde{k}\right)$ to annihilate electrons for $k_+ > e A_-(x_+)$ and to
create positrons for $k_+ < e A_-(x_+)$.

It remains to show that $\Psi\left(x_+,k_+,\widetilde{k}\right)$ is actually an
eigenstate of $-i\partial_+$. Since the first term of (\ref{eq:Psik}) obviously
has this property our task reduces to taking the distributional limit of the
delta sequence $\Delta(k_+ - e A_-;\epsilon)$ in the second term. We shall do
this under the assumption that $k_+$ is well separated from the singular points
at $k_+ = 0$ and at $k_+ = e A_-(x_+)$. Two pieces of notation we shall find
useful are the inverse vector potential $X(k_+)$,
\begin{equation}
k_+ = e A_-(X(k_+)) \; ,
\end{equation}
and the dimensionless ratio of $\widetilde{\omega}^2$ to ($-2 e$ times) the
electric field,
\begin{equation}
\lambda\left(k_+,\widetilde{k}\right) \equiv {\widetilde{\omega}^2 \over 2
e A'_-(X(k_+))} \; . \label{eq:lambda}
\end{equation}

The first step is to change variables from $u$ to $z=[k_+ -e A_-(u)]/\epsilon$, 
\begin{equation}
- \int_L^U dz {i e^{-i (z+i)} \over z + i}
{\cal E}\left(X(k_+ - \epsilon z),x_+;k_+,\widetilde{k}\right) {\Phi_{\infty}
\left(X(k_+ - \epsilon z),\widetilde{k}\right) \over e A_-'(X(k_+ - \epsilon 
z))}\; ,
\label{eq:step1}
\end{equation}
where the upper and lower limits are,
\begin{equation}
U \equiv {k_+ \over \epsilon} \qquad , \qquad L \equiv {k_+ - e A_-(x_+) \over
\epsilon} \; .
\end{equation}
As $\epsilon$ approaches zero they go to positive and negative infinity,
respectively, for $k_+$ in the range $0 < k_+ < e A_-(x_+)$. This is the only
case in which one gets a nonzero result.

We can absorb the Jacobian in (\ref{eq:step1}) by defining a new fundamental 
field,
\begin{equation}
\Xi_{\infty}\left(k_+,\widetilde{k}\right)\equiv \sqrt{2 \pi \over \lambda(k_+,
\widetilde{k})} {\Phi_{\infty}(X(k_+),\widetilde{k}) \over e A_-'(X(k_+))} \; ,
\end{equation}
This brings us to the form,
\begin{equation}
- \int_{L}^{U} {dz \over 2 \pi} {i e^{-i (z+i)} \over z + i} {\cal E}\left(X(
k_+ - \epsilon z),x_+;k_+,\widetilde{k}\right) \sqrt{2 \pi \lambda} \; 
\Xi_{\infty}\left(k_+ - \epsilon z,\widetilde{k}\right) \; . \label{eq:step2}
\end{equation}
Note from (\ref{eq:Phicom}) that the anti-commutator of $\Xi_{\infty}\left(k_+,
\widetilde{k}\right)$ with its adjoint is the same as that of $\Xi_0$ with 
$\Xi_0^{\dagger}$,
\begin{equation}
\left\{\Xi_{\infty}(k_+,\widetilde{k}),\Xi_{\infty}^{\dagger}(q_+,\widetilde{
q})\right\} = \frac1{\sqrt{2}} P_+ (2 \pi)^3 \delta(k_+ - q_+) \delta^2(
\widetilde{k} - \widetilde{q}) \; . \label{eq:Xicom} 
\end{equation}

Now consider the mode function in expression (\ref{eq:step2}),
\begin{equation}
{\cal E}\left(X(k_+ - \epsilon z),x_+;k_+,\widetilde{k}\right) = \exp\left[-
\frac{i}2 \widetilde{\omega}^2 \int_{X(k_+ - \epsilon z)}^{x_+} {du_1 \over k_+
- e A_-(u_1) + i \epsilon}\right] \; .
\end{equation}
For $z < 0$ the lower limit of the integral is a little below the singular
point where the real part of the denominator vanishes. For $z > 0$ the lower 
limit is a little above this point. Straddling the singular point like this 
leads to great sensitivity with respect to $z$, even as $\epsilon$ goes to 
zero. To isolate this $z$ dependence we factor the mode function,
\begin{eqnarray}
\lefteqn{{\cal E}\left(X(k_+ - \epsilon z),x_+;k_+,\widetilde{k}\right) =}
\nonumber \\
& & {\cal E}\left(X(k_+ - \epsilon z),X(k_+);k_+,\widetilde{k}\right) \times 
{\cal E}\left(X(k_+),x_+;k_+,\widetilde{k} \right) \; . \qquad
\end{eqnarray}
The second factor is independent of $z$ and can be pulled outside the integral.
We can also take $\epsilon$ to zero in $\lambda\left(k_+ - \epsilon z,
\widetilde{k}\right)$ and in $\Xi_{\infty}\left(k_+ - \epsilon z,\widetilde{k}
\right)$.

Taking the small $\epsilon$ limit of the first factor requires care. We first 
change variables in the exponent from $u_1$ to $y \equiv [k_+ - e A_-(u_1)]/
\epsilon$ and then expand the Jacobian for small $\epsilon$,
\begin{eqnarray}
\lefteqn{-\frac{i}2 \widetilde{\omega}^2 \int_{X(k_+ - \epsilon z)}^{X(k_+)} 
{du_1 \over k_+ - e A_-(u_1) + i \epsilon}} \nonumber \\
& & \qquad \qquad = -i \lambda\left(k_+,\widetilde{k}\right) \int_0^z {dy 
\over y + i} \times {A_-'(X(k_+)) \over A_-'\left(X(k_+) - \epsilon y\right)} 
\; , \\
& & \qquad \qquad = -i \lambda\left(k_+,\widetilde{k}\right) \ln(z+i) - 
\frac{\pi}2 \lambda\left(k_+,\widetilde{k}\right) + O(\epsilon) \; .
\end{eqnarray}
Dropping the terms which vanish with $\epsilon$ and putting everything together
gives,
\begin{equation}
- \theta(k_+) \theta(e A_-(x_+) - k_+) {\cal E}[A_-]\left(X(k_+),x_+;k_+,
\widetilde{k}\right) \sqrt{2\pi\lambda} \; \gamma(\lambda) \; \Xi_{\infty}
\left(k_+,\widetilde{k}\right) \; , \label{eq:new2}
\end{equation}
where
\begin{equation}
\gamma(\lambda) \equiv e^{-\frac{\pi}2 \lambda} \int_{-\infty}^{\infty} {dz
\over 2\pi} {i e^{-i (z+i)} \over z + i} e^{-i \lambda \ln(z+i)} \; .
\end{equation}

Substituting (\ref{eq:new2}) into (\ref{eq:Psik}) results in the following 
for $\Psi\left(x_+,k_+,\widetilde{k}\right)$:
\begin{eqnarray}
\lefteqn{\Psi\left(x_+,k_+,\widetilde{k}\right) \longrightarrow {\cal 
E}[A_-]\left(0,x_+;k_+,\widetilde{k}\right) \Xi_0\left(k_+,\widetilde{k}\right)}
\nonumber \\
& & - \theta(k_+) \theta(e A_- -k_+) {\cal E}[A_-]\left(X,x_+;k_+,\widetilde{k}
\right) \sqrt{2\pi\lambda} \; \gamma(\lambda) \; \Xi_{\infty}\left(k_+,
\widetilde{k}\right) \; . \label{eq:newPsi}
\end{eqnarray}
We mention again that this is only valid for modes which are well separated 
from the singular points at $k_+ = 0$ and $k_+ = e A_-(x_+)$. If one wishes to
study the behavior of modes which are arbitrarily near either point there is no
alternative to taking a new distributional limit for the delta sequence in
(\ref{eq:Psik}).

Since ${\cal E}[A_-]\left(X(k_+),x_+;k_+,\widetilde{k}\right)$ has the same 
$-i\partial_+$ eigenvalue (\ref{eq:p-}) as the first mode function, $\Psi\left(
x_+,k_+,\widetilde{k}\right)$ is indeed an eigenfunction of $-i \partial_+$. 
This means that it carries a definite energy,
\begin{equation}
\left[H(x_+),\Psi\left(x_+,k_+,\widetilde{k}\right)\right]= {-\widetilde{\omega
}^2/2 \over k_+ - e A_-(x_+)} \Psi\left(x_+,k_+,\widetilde{k}\right) \; .
\label{eq:energy}
\end{equation}
That has implications for the fundamental operators from which it is 
constructed, and for the state upon which they act. The latter is supposed to 
be ``empty'' at $x_+ = 0$. At that instant (\ref{eq:Psik}) implies,
\begin{equation}
\Psi\left(0,k_+,\widetilde{k}\right) = \Xi_0\left(k_+,\widetilde{k}\right) \; .
\end{equation}
Since the potential vanishes at $x_+ = 0$ we can see from (\ref{eq:energy})
that the modes with $k_+ > 0$ carry negative energy while those with $k_+ < 0$
carry positive energy. It follows that the state should obey,
\begin{equation}
\Xi_0\left(k_+,\widetilde{k}\right) \vert \Omega \rangle = 0 = 
\Xi_0^{\dagger}\left(-k_+,-\widetilde{k}\right) \vert \Omega \rangle \qquad 
\forall \; k_+ > 0 \; . \label{eq:state1}
\end{equation}
The $\Xi_{\infty}\left(k_+,\widetilde{k}\right)$ operators (or, equivalently, 
the $\Phi_{\infty}\left(u,\widetilde{k}\right)$ operators) are not present at 
$x_+ = 0$. However, when they do appear --- for $0 < k_+ < e A_-(x_+)$ --- it 
is always with positive energy. It is therefore natural to regard them as 
creators and to define the state to be annihilated by their adjoints,
\begin{equation}
\Xi_{\infty}^{\dagger}\left(k_+,\widetilde{k}\right) \vert \Omega \rangle = 0 
\quad\forall \; k_+ >0 \qquad \Longleftrightarrow \qquad\Phi_{\infty}^{\dagger}
\left(u,\widetilde{k}\right) \vert \Omega \rangle = 0 \quad \forall \; u > 0 
\; . \label{eq:state2}
\end{equation}
What this seems to mean physically is that we allow no particles to enter the
manifold from the surface at $x_- = -\infty$.

Now consider what happens as the system evolves in $x_+$. Under the assumption 
that $e A_-(x_+)$ is an increasing function of $x_+$, modes with $k_+ < 0$ 
begin as positron creation operators and remain that way, although their 
kinetic momenta increase according to the relation, $p_+(x_+) = -k_+ + e 
A_-(x_+)$. The associated mode functions begin as unity and retain unit 
magnitude in the limit that $\epsilon$ vanishes. For $k_+ > 0$ the picture is
more complicated. These modes begin as electron annihilation operators, also 
with mode functions of unit magnitude. However, when $x_+ = X(k_+)$ the energy 
each mode carries passes from $-\infty$ to $+\infty$ and we must regard the
mode as creating a positron. It is not possible to follow this process using 
(\ref{eq:newPsi}) because that expression was derived under the assumption that
the mode was not arbitrarily close to singularity. But we {\it can} use 
(\ref{eq:newPsi}) a little before and a little after singularity. Before 
singularity $\Psi\left(x_+,k_+,\widetilde{k}\right)$ consists of only the term 
proportional to $\Xi_0\left(k_+,\widetilde{k}\right)$, and it has unit 
magnitude. After singularity the magnitude of this term has dropped to $e^{-\pi
\lambda(k_+,\widetilde{k})}$, and the term proportional to $\Xi_{\infty}\left(
k_+,\widetilde{k}\right)$ has appeared. Let us pause at this point to evaluate 
the function $\gamma(\lambda)$ in order to show that the $\Xi_{\infty}\left(
k_+,\widetilde{k}\right)$ term acquires the missing amplitude.

Evaluating $\gamma(\lambda)$ is complicated by the branch cut of the integrand.
However, when $\lambda = -i n$ the integrand is meromorphic and elementary 
methods give $\gamma(-i n) = 1/n!$. By partial integration one can also derive 
the recursion relation $\gamma(\lambda) = (1 + i \lambda) \gamma(\lambda - i)$.
These results together imply that we are dealing with an inverse gamma 
function,
\begin{equation}
\gamma(\lambda) = {1 \over \Gamma(1 + i\lambda)} \; .
\end{equation}
Its magnitude follows from a result of Lobachevskiy, \cite{Grad}
\begin{equation}
{1 \over \Gamma(1 + i\lambda) \Gamma(1 - i\lambda)} = {e^{\pi \lambda} - e^{-
\pi \lambda} \over 2 \pi \lambda} \; . \label{eq:Loba}
\end{equation}
As previously noted, the magnitude of the first mode function ${\cal E}\left(0,
x_+;k_+,\widetilde{k}\right)$ is $e^{-\pi\lambda}$ following the singularity. 
Because the integral in the exponent of the second mode function ${\cal E}
\left(X(k_+),x_+;k_+,\widetilde{k}\right)$ begins precisely at the singularity,
the magnitude of the second mode function is $e^{-\frac{\pi}2 \lambda}$. 
Putting everything together gives the following result for the magnitude of the
various terms multiplying $\Xi_{\infty}\left(k_+,\widetilde{k}\right)$:
\begin{equation}
\left\Vert {\sqrt{2\pi\lambda} \over \Gamma(1 + i \lambda)} {\cal E}\left(
X(k_+),x_+;k_+,\widetilde{k}\right) \right\Vert = \sqrt{1 - e^{-2 \pi \lambda}}
\; . \label{eq:norm}
\end{equation}
Since $\Xi_0\left(k_+,\widetilde{k}\right)$ and $\Xi_{\infty}\left(k_+,
\widetilde{k}\right)$ are independent and canonically normalized operators this
is precisely the correct factor for $\Psi\left(x_+,k_+,\widetilde{k}\right)$ to
retain unit magnitude after singularity.

Heisenberg states can not change but our interpretation of them can. Before 
singularity $\Psi\left(x_+,k_+,\widetilde{k}\right)$ is proportional to 
$\Xi_0\left(k_+,\widetilde{k}\right)$, which annihilates $\vert\Omega \rangle$.
Since $\Psi\left(x_+,k_+,\widetilde{k}\right)$ is an electron annihilation 
operator before singularity this means that both electron spin states with 
$p_+ = k_+ - e A_-(x_+)$ and $\widetilde{p} = \widetilde{k}$ are empty. After 
singularity $\Psi\left(x_+,k_+,\widetilde{k}\right)$ must be a positron 
creation operator because it carries positive charge and energy. If $\Psi\left(
x_+,k_+,\widetilde{k}\right)$ were still proportional to $\Xi_0\left(k_+,
\widetilde{k}\right)$ it would annihilate $\vert \Omega \rangle$ and we should 
have to conclude that both positron spin states with $p_+ = -k_+ + e A_-(x_+)$ 
and $\widetilde{p} = -\widetilde{k}$ had been filled with unit probability. To 
see what actually happens pick the positron spin created by the $i$-th spinor 
component of $\Psi\left(x_+,k_+,\widetilde{k}\right)$ and note that any state
can be written as the sum of a state containing this particle and a state which
does not contain it,
\begin{equation}
\left\vert {\mbox{} \over \mbox{}} \Omega \right\rangle = \sqrt{{\rm Prob}(k_+,
\widetilde{k})} \left\vert {\mbox{} \over \mbox{}} {\rm Full} \right\rangle +
\sqrt{1 - {\rm Prob}(k_+,\widetilde{k})} \left\vert {\mbox{} \over \mbox{}}
{\rm Empty} \right\rangle \; .
\end{equation}
Now act with $2^{1/4} \Psi_i\left(x_+,k_+,\widetilde{k}\right)$ and make 
sequential use of its expansion in terms of $\Xi_0$ and $\Xi_{\infty}$ and the 
fact that it fills the one particle state with unit amplitude,
\begin{eqnarray}
2^{1/4} \Psi_i\left(x_+,k_+,\widetilde{k}\right) \left\vert {\mbox{} \over 
\mbox{}} \Omega \right\rangle & = & {2^{1/4} \sqrt{2 \pi \lambda} \over 
\Gamma(1 + i \lambda)} {\cal E}\left(X,x_+;k_+,\widetilde{k}\right) \Xi_{
\infty \; i} \left\vert {\mbox{} \over \mbox{}} \Omega \right\rangle \; , \\
& = & \sqrt{1 - {\rm Prob}(k_+,\widetilde{k})} \left\vert {\mbox{} \over 
\mbox{}} {\rm Full} \right\rangle \; .
\end{eqnarray}
Use of the anti-commutation relations to compute the norm and comparison with 
(\ref{eq:norm}) shows that the probability for the state to contain a positron 
of this spin is ${\rm Prob}(k_+,\widetilde{k}) = e^{-2 \pi \lambda(k_+,
\widetilde{k})}$. 

Note that we do not see the electron of the electron-positron pair. This is 
because electrons and positrons are both created with $p_+ \sim 0^+$ on the 
lightcone. As explained in Section 2, the positrons accelerate in the $+ z$ 
direction to $p_+ \rightarrow + \infty$, and eventually move parallel to the 
$x_+$ axis. But the electrons accelerate in the $-z$ direction to $p_+ = 0$ and 
therefore leave the manifold moving parallel to the $x_-$ axis immediately 
after creation. We {\it will} see their contribution to the $J_-$ current in 
Section 5.

The picture we have just developed of particle production on the lightcone is
probably the most complete we shall ever have of this otherwise obscure 
phenomenon. To illustrate the power it confers we shall compute the rate per 
unit volume of particle production. For $x_+ > 0$ all modes with $0 < k_+ < e 
A_-(x_+)$ will have passed through singularity, so the probability for the 
entire state to still be in vacuum at this instant is,
\begin{eqnarray}
\lefteqn{P_{\rm vac}(x_+) = \prod_{0 < k_+ < e A_-} \prod_{\widetilde{k}} 
\left(1 - e^{-2 \pi \lambda(k_+,\widetilde{k})}\right)^2 \; ,} \\
& = & \exp\left[V_- \int_0^{e A_-(x_+)} {d k_+ \over 2 \pi} \widetilde{V} \int
{d^2\widetilde{k} \over (2\pi)^2} 2 \ln\left(1 - e^{-2 \pi \lambda(k_+,
\widetilde{k})}\right)\right] \; , \qquad \\
& = & \exp\left[- V_- \widetilde{V} \int_0^{e A_-(x_+)} dk_+ {e A'_-(X(k_+))
\over 4 \pi^3} \sum_{n=1}^{\infty} \frac1{n^2} e^{- \frac{n \pi m^2}{e A'_-(
X)}}\right] \; , \qquad
\end{eqnarray}
where $V_-$ and $\widetilde{V}$ are the volumes of $x_-$ and $\widetilde{x}$
respectively. The rate of production per 4-volume is minus the logarithmic 
derivative of this probability,
\begin{equation}
- {\partial \ln\left[P_{\rm vac}(x_+)\right] \over \partial x_+ \partial V_-
\partial \widetilde{V}} = 
{e A'_-(x_+)^2 \over 4 \pi^3} \sum_{n=1}^{\infty} \frac1{n^2} e^{- \frac{n 
\pi m^2}{e A'_-(x_+)}} \; .
\end{equation}
Note that we do not need to work asymptotically, like Schwinger 
\cite{Schwinger}, nor do we require an {\it ad hoc} interpretation for the 
momentum integral, like Kluger et al. \cite{Kluger2}. It is also significant 
that our result applies for any monotonically decreasing function $A_-(x_+)$. 
It would not be difficult to remove even this restriction.

\section{Back-reaction on the lightcone}

The $\pm$ current operators are nominally $\sqrt{2} e \psi_{\pm}^{\dagger}
\psi_{\pm}$. To enforce invariance under charge conjugation we take one half of
the commutator of the two field operators. To deal with the singularity of 
coincident operators we shall point split in the $x_+$ direction. Since the 
$A_+ = 0$ this procedure is gauge invariant. Since point splitting {\it 
does} break Hermiticity, we shall take the real part,
\begin{eqnarray}
J_{\pm}\left(x_+,x_-,\widetilde{x}\right) & \equiv & {e \over \sqrt{2}} \lim_{
{\Delta x}_+ \rightarrow 0} {\rm Re}\left\{ \psi_{\pm}^{\dagger}\left(x_+,x_-,
\widetilde{x}\right) \psi_{\pm}\left(x_+ + {\Delta x}_+,x_-,\widetilde{x}
\right) \right. \nonumber \\
& & \left.- {\rm Tr}\left[\psi_{\pm}\left(x_+ + {\Delta x}_+,x_-,\widetilde{x}
\right) \psi_{\pm}^{\dagger}\left(x_+,x_-,\widetilde{x}\right)\right]\right\} 
\; .
\end{eqnarray}

To compute the expectation value of $J_+$ it is sufficient to use the 
simplified expansion (\ref{eq:newPsi}) derived in the last section:
\begin{eqnarray}
\lefteqn{\psi_+\left(x_+ + {\Delta x}_+,x_-,\widetilde{x}\right) 
\longrightarrow} \nonumber \\
& & \int_{-\infty}^{\infty} {dk_+ \over 2\pi} e^{-i k_+ x_-} \int {d^2
\widetilde{k} \over (2\pi)^2} e^{i \widetilde{k} \cdot \widetilde{x}} \left\{ 
{\mbox{} \over \mbox{}} {\cal E}\left(0,x_+ + {\Delta x}_+;k_+,\widetilde{k}
\right) \Xi_0\left(k_+,\widetilde{k} \right) \right. \nonumber \\
& & \qquad - \theta(k_+) \theta\left(e A_-(x_+ + {\Delta x}_+) - k_+\right) 
{\sqrt{2 \pi \lambda} \over \Gamma(1 + i \lambda)} \nonumber \\
& & \left. {\mbox{} \over \mbox{}} \qquad \qquad \times {\cal E}\left(X(k_+),
x_+ + {\Delta x}_+;k_+,\widetilde{k}\right) \Xi_{\infty}\left(k_+,\widetilde{k}
\right)\right\} \; , \\
\lefteqn{\psi_+^{\dagger}\left(x_+,x_-,\widetilde{x}\right) \longrightarrow}
\nonumber \\
& & \int_{-\infty}^{\infty} {dk_+ \over 2\pi} e^{i k_+ x_-} \int {d^2
\widetilde{k} \over (2\pi)^2} e^{-i \widetilde{k} \cdot \widetilde{x}} \left\{ 
{\mbox{} \over \mbox{}} {\cal E}^*\left(0,x_+;k_+,\widetilde{k}\right) \Xi_0^{
\dagger}\left(k_+,\widetilde{k} \right) \right. \nonumber \\
& & \left. - \theta(k_+) \theta\left(e A_-(x_+) - k_+\right) {\sqrt{2 \pi 
\lambda} \over \Gamma(1 - i \lambda)} {\cal E}^*\left(X,x_+;k_+,\widetilde{k}
\right) \Xi_{\infty}^{\dagger}\left(k_+,\widetilde{k}\right) \right\} \; . 
\qquad
\end{eqnarray}
We note also two important identities concerning the mode function ${\cal E}$:
\begin{eqnarray}
\lefteqn{{\cal E}^*\left(0,x_+;k_+\widetilde{k}\right) {\cal E}\left(0,x_+ +
{\Delta x}_+;k_+,\widetilde{k}\right)} \nonumber \\
& & \qquad \qquad = e^{-2\pi \lambda \theta(k_+) \theta(eA_- - k_+)} 
{\cal E}\left(x_+,x_+ + {\Delta x}_+;k_+,\widetilde{k}\right) \; , \\
\lefteqn{{\cal E}^*\left(X,x_+;k_+\widetilde{k}\right) {\cal E}\left(X,x_+ +
{\Delta x}_+;k_+,\widetilde{k}\right)} \nonumber \\
& & \qquad \qquad = e^{-\pi \lambda} {\cal E}\left(x_+,x_+ + {\Delta x}_+;k_+,
\widetilde{k}\right) \; .
\end{eqnarray}
Combining these relations with the conditions (\ref{eq:state1}-\ref{eq:state2})
which define the state and the anti-commutation relations (\ref{eq:Psicom})
and (\ref{eq:Xicom}) we obtain the following result for the expectation value 
of $J_+$:
\begin{eqnarray}
\lefteqn{\left\langle \Omega \left\vert J_+\left(x_+,x_-,\widetilde{x}\right)
\right\vert \Omega \right\rangle} \nonumber \\
& = & e\lim_{{\Delta x}_+ \rightarrow 0^+}\int {d^2\widetilde{k} \over(2\pi)^2}
{\rm Re}\left\{ \int_{-\infty}^0 {dk_+ \over 2\pi} + \int_0^{eA_-} {dk_+ \over 
2\pi} \left[1-e^{-2 \pi \lambda(k_+,\widetilde{k})}\right] \right. \nonumber \\
& & \left. - \int_0^{eA_-} {dk_+ \over 2\pi} e^{-2\pi \lambda(k_+,
\widetilde{k})} - \int_{eA_-}^{\infty} {dk_+ \over 2\pi} \right\} {\cal E}
\left(x_+,x_+ + {\Delta x_+};k_+,\widetilde{k}\right) \; , \\ \label{eq:firstJ}
& = & -2 e \int_0^{e A_-} {dk_+ \over 2\pi} \int {d^2\widetilde{k} \over (2
\pi)^2} \; e^{-2\pi \lambda(k_+,\widetilde{k})} \nonumber \\
& & + e \lim_{{\Delta x}_+ \rightarrow 0^+} {\rm Re}\left\{\int_0^{\infty} 
{dq \over 2\pi} \int {d^2\widetilde{k} \over (2\pi)^2} \left[{\cal E}\left(x_+,
x_+ {\Delta x}_+;-q + eA_-,\widetilde{k}\right) \right. \right. \nonumber \\
& & \qquad \qquad \left. {\mbox{} \over \mbox{}} \left. - {\cal E}\left(x_+,x_+ 
+ {\Delta x}_+;q + e A_-,\widetilde{k}\right)\right]\right\} \; . 
\label{eq:intJ+}
\end{eqnarray}

The final term in (\ref{eq:intJ+}) vanishes. To see this first perform the 
integration over $\widetilde{k}$,
\begin{equation}
\int {d^2\widetilde{k} \over (2\pi)^2} {\cal E}\left(x_+,x_+ + {\Delta x}_+; 
\pm q + e A_-,\widetilde{k}\right) = {-i \over 2 \pi} {e^{-\frac{i}2 m^2 
I({\Delta x}_+,\pm q)} \over I({\Delta x}_+,\pm q)} \; ,
\end{equation}
where we define
\begin{equation}
I({\Delta x}_+,\pm q) \equiv \int_{x_+}^{x_+ + {\Delta x}_+} {du \over \pm
q + e A_-(x_+) - e A_-(u) + i \epsilon} \; . \label{eq:Iint}
\end{equation}
This brings the final term in (\ref{eq:intJ+}) to the form,
\begin{equation}
{e \over 4 \pi^2} \lim_{{\Delta x}_+ \rightarrow 0^+} {\rm Im}\left\{ \int_0^{
\infty} dq \left[ {e^{-\frac{i}2 m^2 I({\Delta x}_+,-q)} \over I({\Delta x}_+,
-q)} - {e^{-\frac{i}2 m^2 I({\Delta x}_+,+q)} \over I({\Delta x}_+,+q)} \right]
\right\} \; . \label{eq:2ndJ+}
\end{equation}
The function $I({\Delta x}_+,\pm q)$ can be expanded in powers of the splitting
parameter ${\Delta x}_+$,
\begin{eqnarray}
I({\Delta x}_+,\pm q) & = & {{\Delta x}_+ \over \pm q + i \epsilon} \left\{1 + 
\frac12 \left[{e A_-' {\Delta x}_+ \over \pm q + i \epsilon}\right] + \frac16 
\left[{e A_-^{\prime \prime} {\Delta x}_+^2 \over \pm q + i \epsilon} \right] 
\right.  \nonumber \\
& & \qquad \qquad \qquad \left. + \frac13 \left[{e A_-' {\Delta x}_+ \over 
\pm q + i \epsilon}\right]^2 + O\left({\Delta x}_+^3\right) \right\} \; .
\label{eq:Iexp}
\end{eqnarray}
Since $I({\Delta x}_+,\pm q)$ goes to zero with ${\Delta x}_+$, and for large 
$q$, we can expand the exponentials of (\ref{eq:2ndJ+}) inside the $q$ 
integration. When this is done it is easy to see that every term vanishes 
either in taking the imaginary part or in taking ${\Delta x}_+$ to zero,
\begin{eqnarray}
\lefteqn{\lim_{{\Delta x}_+ \rightarrow 0^+} {\rm Im}\left\{ {e^{-\frac{i}2 m^2
I({\Delta x}_+,-q)} \over I({\Delta x}_+,-q)} - {e^{-\frac{i}2 m^2 I({\Delta 
x}_+,+q)} \over I({\Delta x}_+,+q)} \right\}} \nonumber \\
& = & \lim_{{\Delta x}_+ \rightarrow 0^+} {\rm Im}\left\{ {1 \over I({\Delta 
x}_+,-q)} - \frac{i}2 m^2 + \dots - {1 \over I({\Delta x}_+, +q)} + \dots
\right\} \; , \\
& = & \lim_{{\Delta x}_+ \rightarrow 0^+} {\rm Im}\left\{ {-q + i\epsilon \over
{\Delta x}_+} - \frac12 e A_-' + \dots - {(q + i\epsilon) \over {\Delta x}_+} +
\frac12 e A_-' + \dots \right\} \; , \\
& = & 0 \; .
\end{eqnarray}

$J_+$ gives the charge density on surfaces of constant $x_+$ and we have seen
that its expectation value is,
\begin{eqnarray}
\left\langle \Omega \left\vert J_+\left(x_+,x_-,\widetilde{x}\right) 
\right\vert \Omega \right\rangle & = & -2 e \int_0^{e A_-} {dk_+ \over 2 \pi}
\int {d^2\widetilde{k} \over (2\pi)^2} e^{-2\pi \lambda(k_+,\widetilde{k})}
\; , \label {eq:firstform} \\
& = & -\frac{e}{\pi} \int_0^{x_+} du \left[{e A_-'(u) \over 2 \pi}\right]^2 
\exp\left[- {\pi m^2 \over e A_-'(u)}\right] \; . \qquad 
\end{eqnarray}
The first form (\ref{eq:firstform}) is actually the simplest to understand
physically. It says that the charge density accumulates each of the two 
positron spin states with probability $e^{-2\pi \lambda}$ as the mode with 
canonical momenta $k_+$ and $\widetilde{k}$ passes through singularity. As 
noted before, the electron partners in the pair creation event accelerate to 
the speed of light in the $-z$ direction and leave the manifold moving parallel
to the $x_-$ axis. It might seem that since the manifold becomes charged the 
vector potential must depend upon $x_-$, and we have therefore not solved the 
problem for a sufficiently general class of potentials to include the actual 
back-reacted solution. However, we shall see that the response from $J_-$ is 
actually infinite and infinitely fast --- precisely {\it because} the electrons
have exited by reaching the speed of light. This means that back-reaction 
drives the actual potential to zero infinitely fast, before $J_+$ can become 
nonzero.

Evaluating the expectation value of $J_-(x_+,x_-,\widetilde{x})$ is complicated
because the result must diverge as $\epsilon$ approaches zero. To see this note
that since the expectation value of $\widetilde{J}(x_+,x_-,\widetilde{x})$ 
vanishes, current conservation and our result (\ref{eq:firstform}) for the
expectation value of $J_+(x_+,x_-,\widetilde{x})$ imply,
\begin{eqnarray}
\partial_- \left\langle \Omega \left\vert J_-\left(x_+,x_-,\widetilde{x}\right)
\right\vert \Omega \right\rangle & = & -\partial_+ \left\langle \Omega 
\left\vert J_+\left(x_+,x_-,\widetilde{x}\right) \right\vert \Omega 
\right\rangle \; , \\
& = & \frac1{\pi} e^2 A_-'(x_+) \int {d^2\widetilde{k} \over (2\pi)^2} e^{-2 
\pi \lambda(eA_-,\widetilde{k})} \; .
\end{eqnarray}
Integration from the lower limit of our $\epsilon$-regulated range of $x_-$
gives,
\begin{eqnarray}
\left\langle \Omega \left\vert J_-\left(x_+,x_-,\widetilde{x}\right) 
\right\vert \Omega \right\rangle & = & \left\langle \Omega \left\vert J_-\left(
x_+,-\epsilon^{-1},\widetilde{x}\right) \right\vert \Omega \right\rangle 
\nonumber \\
& & + \left(x_- + \frac1{\epsilon}\right) \frac1{\pi} e^2 A_-'(x_+) \int 
{d^2\widetilde{k} \over (2\pi)^2} e^{-2 \pi \lambda(eA_-,\widetilde{k})} \; .
\quad
\end{eqnarray}
The final term has a simple physical interpretation. $J_-$ is a charge flux so
it must register the newly created electrons which rush off the manifold 
parallel to the $x_-$ axis. (Because they are moving in the $-z$ direction at 
the speed of light.) The rate at which this charge is created, per unit volume 
in $x_-$ and $\widetilde{x}$, is just $-\partial_+ J_+$. An electron created at
position $(x_+,x_-,\widetilde{x})$ must pass through all points $(x_+,y_+,
\widetilde{x})$, for $y_- > x_-$, on its way off the manifold. So the net
electronic flux through any point $x_-$ is the integral of $-\partial_+ 
J_+(x_+,y_-,\widetilde{x})$ over all points $y_- < x_-$. We have cut the lower
limit off at $-1/\epsilon$, so the electronic contribution to the expectation
value of $J_-$ must diverge as $\epsilon$ goes to zero.

Although there is a good physical reason for it, the fact that the expectation
value of $J_-$ diverges like $1/\epsilon$ means that we must use special care
in evaluating distributional limits which involve $\epsilon$. For example, the
field equations can be inverted to give $\widetilde{\psi}_-$ in terms of
$\widetilde{\psi}_+$,
\begin{equation}
\widetilde{\psi}_-\left(x_+,x_-,\widetilde{k}\right) = \left({m - \widetilde{
\gamma} \cdot \widetilde{k} \over \widetilde{\omega}^2}\right) \gamma_+ i
\partial_+ \widetilde{\psi}_+\left(x_+,x_-,\widetilde{k}\right) \; .
\end{equation}
However, we cannot simply substitute the $x_+$ derivative of expression 
(\ref{eq:newPsi}) because the distributional limit in the second term of that 
formula was computed assuming that $k_+$ is separated from zero and $e 
A_-(x_+)$. When $\partial_+$ acts upon the second $\theta$-function in 
(\ref{eq:newPsi}) it gives a $\delta$-function which invalidates that 
assumption by setting $k_+ = e A_-(x_+)$. One can tell from the ultralocality 
of this term at $k_+= e A_-(x_+)$ that it is responsible for the electronic 
contribution computed above. Rather than forcing everything through from the 
cumbersome, initial expressions we shall just compute the expectation value of 
$J_-$ without this term and then compensate by adding in the electron current 
found above.

Our computational shortcut amounts to making the following replacements:
\begin{eqnarray}
\lefteqn{\psi_-\left(x_+ + {\Delta x}_+,x_-,\widetilde{x}\right) 
\longrightarrow } \nonumber \\
& & \int_{-\infty}^{\infty} {dk_+ \over 2\pi} e^{-i k_+ x_-} \int 
{d^2 \widetilde{k} \over (2\pi)^2} e^{i \widetilde{k}\cdot \widetilde{x}} 
{\frac12 (m - \widetilde{\gamma} \cdot \widetilde{k}) \gamma_+ \Psi\left(x_+ + 
{\Delta x}_+,k_+,\widetilde{k}\right) \over k_+ - e A_-(x_+ + {\Delta x}_+) + 
i \epsilon} \; , \\
\lefteqn{\psi_-^{\dagger}\left(x_+,x_-,\widetilde{x}\right) \longrightarrow }
\nonumber \\
& & \; \int_{-\infty}^{\infty} {dk_+ \over 2\pi} e^{i k_+ x_-} \int {d^2
\widetilde{k} \over (2\pi)^2} e^{-i \widetilde{k} \cdot \widetilde{x}} 
{\Psi^{\dagger}\left(x_+,k_+,\widetilde{k}\right) \frac12 \gamma_- (m +
\widetilde{\gamma} \cdot \widetilde{k}) \over k_+ - e A_-(x_+) - i \epsilon} 
\; . \quad
\end{eqnarray}
The integrand of this expression for $\psi_-$ is just the same as for $\psi_+$ 
multiplied by a factor $\frac12 (m - \widetilde{\gamma} \cdot \widetilde{k}) 
\gamma_+$ and divided by $p_+ + i \epsilon$ at the appropriate value of $x_+$. 
Hence the contribution to the expectation value of $J_-$ is the same as to 
$J_+$ but with an additional factor of
\begin{equation}
{\frac12 (m - \widetilde{\gamma} \cdot \widetilde{k}) \gamma_+ \over p_+(x_+ +
{\Delta x}_+) + i \epsilon} \times {\frac12 \gamma_- (m + \widetilde{\gamma}
\cdot \widetilde{k}) \over p_+(x_+) - i \epsilon} = {\frac12 \widetilde{
\omega}^2 \; P_- \over [p_+(x_+ + {\Delta x}_+) + i \epsilon] [p_+(x_+) - i
\epsilon]} \; .
\end{equation}
The expectation value of $J_-$, sans its electronic component, can therefore
be obtained by simply including this factor in our previous expression 
(\ref{eq:firstJ}) for the expectation value of $J_+$,
\begin{eqnarray}
\lefteqn{\left\langle \Omega \left\vert J_-\left(x_+,x_-,\widetilde{x}\right) 
\right\vert \Omega \right\rangle - ({\rm electronic\ contribution}) =
e \lim_{{\Delta x}_+ \rightarrow 0^+} } \nonumber \\
& & \int {d^2\widetilde{k} \over 
(2\pi)^2} {\rm Re}\left\{\left[\int_{-\infty}^0 {dk_+ \over 2\pi} + \int_0^{e 
A_-} {dk_+ \over 2 \pi} \left(1 - e^{-2\pi \lambda}\right) - \int_0^{eA_-} 
{dk_+ \over 2\pi} e^{-2\pi \lambda} \right.\right. \nonumber \\
& & \left. \left. - \int_{e A_-}^{\infty} {dk_+ \over 2 \pi} \right] {\frac12 
\widetilde{\omega}^2 {\cal E}\left(x_+,x_+ + {\Delta x}_+;k_+,\widetilde{k}
\right) \over [k_+ - e A_-(x_+ + {\Delta x}_+) + i \epsilon] [k_+ - e A_-(x_+) 
- i \epsilon]} \right\} \; , \\
& & = - 2 e \int_0^{e A_-} {d k_+ \over 2\pi} \int {d^2\widetilde{k} \over (2
\pi)^2} {\frac12 \widetilde{\omega}^2 e^{-2\pi \lambda(k_+,\widetilde{k})} 
\over [k_+ - e A_-(x_+)]^2 + \epsilon^2}  + e \lim_{{\Delta x}_+ \rightarrow 
0^+} {\partial \over \partial {\Delta x}_+} \nonumber \\
& & \times {\rm Re}\left\{ \int_0^{\infty} {dq \over 2\pi} \int {d^2\widetilde{
k} \over (2\pi)^2} \left[-{i e^{-\frac{i}2 \widetilde{\omega}^2 I({\Delta x}_+,
-q)} \over q + i\epsilon} -{i e^{-\frac{i}2 \widetilde{\omega}^2 I({\Delta 
x}_+,+q)} \over q - i\epsilon} \right] \right\} \; , \quad \label{eq:J-}
\end{eqnarray}
where the function $I({\Delta x}_+,q)$ is defined in equation (\ref{eq:Iint}).

The first term of (\ref{eq:J-}) has a simple interpretation as the 
($\epsilon$-regulated) current due to the created positrons. Each of the two
positron spin states is created with probability $e^{-\pi \lambda(k_+,
\widetilde{k})}$, and each contributes a factor of $-e p_-/p_+$ to the current
density. It is simple to perform the integration over $\widetilde{k}$ and to
recast the remaining integration to one over $x_+$,
\begin{eqnarray}
\lefteqn{- 2 e \int_0^{e A_-} {d k_+ \over 2\pi} \int {d^2\widetilde{k} \over 
(2 \pi)^2} {\frac12 \widetilde{\omega}^2 e^{-2\pi \lambda(k_+,\widetilde{k})} 
\over [k_+ - e A_-(x_+)]^2 + \epsilon^2}} \nonumber \\
& = & -{e \over 8 \pi^4} \int_0^{x_+} du [e A_-'(u)]^2 {[\pi m^2 + e A_-'(u)]
e^{{-\pi m^2 \over e A_-'(u)}} \over [e A_-(u) - e A_-(x_+)]^2 + \epsilon^2}
\; . \label{eq:J-1}
\end{eqnarray}

Although the positron current can diverge like $1/\epsilon$ it must vanish at 
$x_+ =0$. This crucial fact distinguishes it from the electron current,
\begin{equation}
\left(x_- + \frac1{\epsilon}\right) \frac1{\pi} e^2 A_-'(x_+) \int {d^2
\widetilde{k} \over (2\pi)^2} e^{-2 \pi \lambda(eA_-,\widetilde{k})} = 
\left(x_- + \frac1{\epsilon}\right) {e^3 A_-^{\prime 2}(x_+) \over 4 \pi^3}
e^{{-\pi m^2 \over e A_-'(x_+)}} \; . \label{eq:J-2}
\end{equation}
Even though the state is initially empty there is no way to prevent particle
production at $x_+ = 0$ because there are modes with $k_+$ arbitrarily close
to zero. The electron current comes entirely from particles which are created
moving with the speed of light at the same instant that the current is being
measured, so it must be present even at $x_+ = 0$. This means that the negative
electron current must initially dominate the positive positron current. Hence
back-reaction acts in the physically sensible direction to reduce the initial
electric field. Since the initial electron current is not only negative 
definite but infinite, as $\epsilon$ goes to zero, back-reaction becomes 
infinitely strong, infinitely fast.

The final term in (\ref{eq:J-}) is the charge renormalization. One sees this
because it contains the logarithmic ultraviolet divergence and because it is 
proportional to the right hand side of the relevant one of Maxwell's equations 
for this background,
\begin{equation}
-A_-^{\prime\prime}(x_+) = \left\langle \Omega \left\vert J_-\left(x_+,x_-,
\widetilde{x}\right) \right\vert \Omega \right\rangle \; . \label{eq:Max}
\end{equation}
We evaluate it by the same strategy as for the analogous (vanishing)  
contribution to $J_+$. First perform the integration over $\widetilde{k}$, 
then expand in powers of the function $I({\Delta x}_+,\pm q)$, expand $I({
\Delta x}_+,\pm q)$ in powers of ${\Delta x}_+$ according to (\ref{eq:Iexp}), 
and finally take the derivative, the real part and the limit inside the 
integration over $q$. The result is,
\begin{eqnarray}
\lefteqn{{e \over 4 \pi^2} \lim_{{\Delta x}_+ \rightarrow 0^+} {\partial \over 
\partial {\Delta x}_+} {\rm Re}\left\{\int_0^{\infty} dq \left[-{e^{-\frac{i}2 
m^2 I({\Delta x}_+,-q)} \over (q +i \epsilon) I} - {e^{-\frac{i}2 m^2 I({\Delta
x}_+,+q)} \over (q - i\epsilon) I} \right] \right\}} \nonumber \\
& = & {e \over 4 \pi^2} \lim_{{\Delta x}_+ \rightarrow 0^+} {\partial \over 
\partial {\Delta x}_+} {\rm Re}\left\{\int_0^{\infty} dq \right. \nonumber \\
& & \left[-{1 \over q +i \epsilon} \left({1 \over I({\Delta x}_+,-q)} - 
\frac{i}2 m^2 - \frac18 m^4 I({\Delta x}_+,-q) + \dots \right) \right. 
\nonumber \\
& & \left. \left. - {1 \over q - i\epsilon} \left({1 \over I({\Delta x
}_+,+q)} - \frac{i}2 m^2 - \frac18 m^4 I({\Delta x}_+,+q) + \dots \right) 
\right] \right\} \; , \qquad \\
& = & {e \over 4 \pi^2} \lim_{{\Delta x}_+ \rightarrow 0^+} {\partial \over 
\partial {\Delta x}_+} {\rm Re}\left\{\int_0^{\infty} dq \right. \nonumber \\
& & \left[{1 \over {\Delta x}_+} + {\frac12 e A_-' \over q + i \epsilon}
+ {\frac16 e A_-^{\prime\prime} {\Delta x}_+ \over q + i \epsilon} - 
{\left[\frac1{12} (e A_-')^2 + \frac18 m^4\right] {\Delta x}_+ \over q^2 + 
\epsilon^2} + \dots \right. \nonumber \\
& - & \left. \left. {1 \over {\Delta x}_+} + {\frac12 e A_-' \over q - i
\epsilon} + {\frac16 e A_-^{\prime\prime} {\Delta x}_+ \over q - i \epsilon} + 
{\left[\frac1{12} (e A_-')^2 + \frac18 m^4\right] {\Delta x}_+ \over q^2 +
\epsilon^2} + \dots \right] \right\} \; , \qquad \\
& = & {e \over 12 \pi^2} e A_-^{\prime\prime}(x_+) \int_0^{\infty} dq {q \over
q^2 + \epsilon^2} \; . \label{eq:J-3}
\end{eqnarray}
Had we computed the expectation value of $J_+$ for a more general class of 
vector potentials depending upon $x_-$ as well as $x_+$, we would have found 
the same term as (\ref{eq:J-3}) but with $A_-^{\prime\prime}(x_+)$ replaced by 
$- \partial_+ \partial_- A_-(x_+,x_-)$.

To get the one loop correction to Maxwell's equation (\ref{eq:Max}) we must
combine the constituents of the expectation value of $J_-$: expressions
(\ref{eq:J-1}), (\ref{eq:J-2}) and (\ref{eq:J-3}). Since (\ref{eq:J-3}) is a
charge renormalization its proper place is on the left hand side of the 
equation. The result is,
\begin{eqnarray}
\lefteqn{-\left[1 + {e^2 \over 12 \pi^2} \int_0^{\infty} dq {q \over q^2 + 
\epsilon^2} \right] A_-^{\prime\prime}(x_+) = \left(x_- + \frac1{\epsilon}
\right) {e^3 A_-^{\prime 2}(x_+) \over 4 \pi^3} e^{{-\pi m^2 \over e 
A_-'(x_+)}}} \qquad \nonumber \\
& & -{e \over 8 \pi} \int_0^{x_+} du \left[{e A_-'(u) \over \pi}\right]^2
{\left[m^2 + {e A_-'(u) \over \pi}\right] e^{{-\pi m^2 \over e A_-'(u)}} \over
[e A_-(u) - e A_-(x_+)]^2 + \epsilon^2} \; . \label{eq:Maxp}
\end{eqnarray}
Now recall from standard QED that the renormalized charge $e_R$ and field 
$A_R(x_+)$ are related to the unrenormalized ones by square roots of the field
strength $Z$,
\begin{equation}
e_R \equiv \sqrt{Z} e \qquad , \qquad A_R(x_+) \equiv {1 \over \sqrt{Z}} 
A_-(x_+) \; .
\end{equation}
Note particularly that $e A_-(x_+) = e_R A_R(x_+)$. Multiplying (\ref{eq:Maxp})
by $\sqrt{Z}$ we obtain,
\begin{eqnarray}
\lefteqn{- \left[Z + {e_R^2 \over 12 \pi^2} \int_0^{\infty} dq {q \over q^2 + 
\epsilon^2} \right] A_R^{\prime\prime}(x_+) = \left(x_- + \frac1{\epsilon}
\right) {e^3 A_-^{\prime 2}(x_+) \over 4 \pi^3} e^{{-\pi m^2 \over e 
A_-'(x_+)}}} \qquad \nonumber \\
& & -{e_R \over 8 \pi} \int_0^{x_+} du \left[{e_R A_R'(u) \over \pi}\right]^2
{\left[m^2 + {e_R A_R'(u) \over \pi}\right] e^{{-\pi m^2 \over e_R A_R'(u)}} 
\over [e_R A_R(u) - e_R A_R(x_+)]^2 + \epsilon^2} \; . \label{eq:Maxpp}
\end{eqnarray}
If we recognize the one loop field strength renormalization as
\begin{equation}
Z = 1 - {e_R^2 \over 12 \pi^2} \int_0^{\infty} dq {q \over q^2 + \epsilon^2}
\; ,
\end{equation}
(up to finite renormalizations) then the equation assumes its standard form,
\begin{eqnarray}
\lefteqn{- A_R^{\prime\prime}(x_+) = \left(x_- + \frac1{\epsilon} \right) 
{e_R^3 A_R^{\prime 2}(x_+) \over 4 \pi^3} e^{{-\pi m^2 \over e_R
A_R'(x_+)}}} \qquad \nonumber \\
& & -{e_R \over 8 \pi} \int_0^{x_+} du \left[{e_R A_R'(u) \over \pi}\right]^2
{\left[m^2 + {e_R A_R'(u) \over \pi}\right] e^{{-\pi m^2 \over e_R A_R'(u)}} 
\over [e_R A_R(u) - e_R A_R(x_+)]^2 + \epsilon^2} \; . \label{eq:Maxppp}
\end{eqnarray}

For small $\epsilon$ (which we must take to zero anyway) the instantaneous
electron current dominates the positron current and the equation becomes {\it 
local},
\begin{equation}
- A_R^{\prime\prime}(x_+) \approx {e_R^3 A_R^{\prime 2}(x_+) \over 4 \pi^3 
\epsilon} e^{{-\pi m^2 \over e_R A_R'(x_+)}} \; . \label{eq:newMax}
\end{equation}
When compared with the sorts of equations one finds for the traditional problem
of evolving from a surface of constant $x^0$ (for example, see Section 3 of 
\cite{Kluger}) expression (\ref{eq:newMax}) is almost unbelievably simple. We
can simplify it further by rescaling both the evolution variable,
\begin{equation}
\tau \equiv \left({e_R m \over 2 \pi}\right)^2 {x_+ \over \epsilon} \; ,
\end{equation}
and the electric field,
\begin{equation}
F(\tau) \equiv {e_R A_R'(x_+) \over \pi m^2} \; .
\end{equation}
The result is a first order, ordinary differential equation,
\begin{equation}
{d \over d\tau} e^{F^{-1}} = 1 \; .
\end{equation}
The solution is straightforward,
\begin{equation}
F(\tau) = {1 \over \ln\left(e^{1/F_0} + \tau\right)} \; .
\end{equation}
Since $\tau$ approaches infinity for any fixed, positive value of $x_+$ our
solution means that back-reaction forces the electric field to zero before any
fixed, positive value of $x_+$. This is as far as the equations can be used
because they were derived under the assumption that $e A_R(x_+)$ is an 
increasing function of $x_+$. Note that our solution also implies the vanishing
of the vector potential before any fixed, positive value of $x_+$. So the
expectation value of $J_+$ is really zero at the physical solution, and there
is no need to consider backgrounds which depend upon $x_-$.

\section{The infinite boost correspondence limit}

The results of the past section have a single unsatisfying feature: the factors
of $1/\epsilon$ in the expectation value of $J_-$ mean that back-reaction on
the lightcone becomes infinitely strong, infinitely fast. This seems to be in
dramatic distinction with what happens for the traditional problem in which the
state is released on a surface of constant $x^0$. There the induced current 
grows smoothly from $x^0 = 0$, and it remains finite for finite $x^0$. The 
purpose of this section is to show that our result is not distinct from the 
traditional one. Rather the problem we have worked out can be viewed as the 
infinite boost limit of the traditional problem, in the same way that lightcone
quantum field theory can always be viewed as the infinite momentum frame 
\cite{Kogut}.

\begin{figure}
\centerline{\epsfig{file=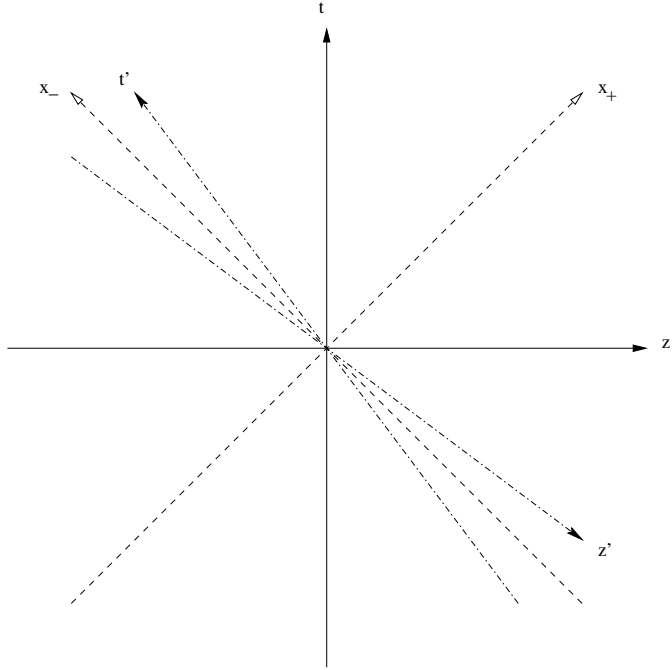,height=3.5in}}
\caption{The various coordinate systems.}
\end{figure}

To fix notation let us consider two inertial frames. The one in which we have 
been working will be denoted the unprimed frame. The primed frame moves with
speed $\beta$ along the minus $z$ axis, so the Lorentz transformation between
the two systems is,
\begin{eqnarray}
t' & = & \gamma (t + \beta z) \; , \\
z' & = & \gamma (z + \beta t) \; ,
\end{eqnarray}
where $\gamma \equiv 1/\sqrt{1 - \beta^2}$. Note that the time coordinate of
the primed frame has the following expression in terms of the lightcone 
coordinates in the unprimed frame,
\begin{equation}
t' = \sqrt{1 + \beta \over 1 - \beta} {x_+ \over \sqrt{2}} + \sqrt{1 - \beta
\over 1 + \beta} {x_- \over \sqrt{2}} \; . \label{eq:tprime}
\end{equation}
The relation between the two frames is shown in Fig.~2.

We wish to compare evolution in $x_+$ in the unprimed frame with evolution in 
primed frame in the limit that $\beta$ approaches one. We assume that the
vector potential and the current density of the primed frame depend only upon
$t'$ and have the form,
\begin{eqnarray}
A_0'(t') = 0 \qquad & , & \qquad A_3'(t') = A(t') \; , \\
J^{\prime 0}(t') = 0 \qquad & , & \qquad J^{\prime 3}(t') = J(t') \; .
\end{eqnarray}
Transforming the vector potential covariantly gives,
\begin{eqnarray}
A_0 & = & \gamma (A_0' + \beta A_3') = \beta \gamma A(t') \; , \\
A_3 & = & \gamma (A_3' + \beta A_0') = \gamma A(t') \; .
\end{eqnarray}
The current density transforms as a contravariant vector to give,
\begin{eqnarray}
J^0 & = & \gamma (J^{\prime 0}-\beta J^{\prime 3})= -\beta \gamma J(t') \; , \\
J^3 & = & \gamma (J^{\prime 3} - \beta J^{\prime 0}) = \gamma J(t') \; .
\end{eqnarray}

The lightcone components $A_{\pm} = (A_0 \pm A_3)/\sqrt{2}$ of the unprimed 
frame vector potential are,
\begin{eqnarray}
A_+(x_+,x_-) & = & {1 \over \sqrt{2}} \sqrt{1 + \beta \over 1 - \beta} A\left(
\sqrt{1 + \beta \over 1 - \beta} {x_+ \over \sqrt{2}} + \sqrt{1 - \beta \over 1
+ \beta} {x_- \over \sqrt{2}}\right) \; ,\\
A_-(x_+,x_-) & = & {-1 \over \sqrt{2}} \sqrt{1 - \beta \over 1 + \beta} A\left(
\sqrt{1 + \beta \over 1 - \beta} {x_+ \over \sqrt{2}} + \sqrt{1 - \beta \over 1
+ \beta} {x_- \over \sqrt{2}}\right) \; .
\end{eqnarray}
We can enforce our $A_+ = 0$ gauge condition with the tranformation,
\begin{equation}
\widehat{A}_{\pm}(x_+,x_-) = A_{\pm}(x_+,x_-) - \partial_{\pm} \int_0^{t'} ds' 
A(s') \; ,
\end{equation}
which gives,
\begin{equation}
\widehat{A}_-(x_+,x_-) = -\sqrt{2} \sqrt{1 - \beta \over 1 + \beta} A\left(
\sqrt{1 + \beta \over 1 - \beta} {x_+ \over \sqrt{2}} + \sqrt{1 - \beta \over 1
+ \beta} {x_- \over \sqrt{2}}\right) \; . \label{eq:vec}
\end{equation}
The (gauge invariant) electric field is,
\begin{equation}
E(x_+,x_-) = -\partial_+ \widehat{A}_-(x_+,x_-) = A'\left(\sqrt{1 + \beta \over
1 - \beta} {x_+ \over \sqrt{2}} + \sqrt{1 - \beta \over 1 + \beta} {x_- \over 
\sqrt{2}}\right) \; . \label{eq:Ez}
\end{equation}
The lightcone components $J_{\pm} = (J^0 \pm J^3)/\sqrt{2}$ of the unprimed 
frame current vector are, 
\begin{eqnarray} 
J_+(x_+,x_-) & = & {1 \over \sqrt{2}} \sqrt{1 - \beta \over 1 + \beta} J\left(
\sqrt{1 + \beta \over 1 - \beta} {x_+ \over \sqrt{2}} + \sqrt{1 - \beta \over 1
+ \beta} {x_- \over \sqrt{2}}\right) \; , \label{eq:J++} \\
J_-(x_+,x_-) & = & {-1 \over \sqrt{2}} \sqrt{1 + \beta \over 1 - \beta} J\left(
\sqrt{1 + \beta \over 1 - \beta} {x_+ \over \sqrt{2}} + \sqrt{1 - \beta \over 1
+ \beta} {x_- \over \sqrt{2}}\right) \; . \label{eq:J--}
\end{eqnarray}

The key relations are (\ref{eq:vec}-\ref{eq:J--}). Let us consider them as 
$\beta$ approaches one, first under the assumption that back-reaction is turned
off. In this case the electric field is constant so the vector potential and 
the current density in the primed frame both grow linearly in $t'$,
\begin{equation}
A(t') = E_0 t' \qquad , \qquad J(t') = J_0 t' \; .
\end{equation}
From relations (\ref{eq:vec}) and (\ref{eq:Ez}) we see that the lightcone 
vector potential is also linear, and the electric field is also constant,
\begin{equation}
\widehat{A}_-(x_+,x_-) \longrightarrow - E_0 x_+ \qquad , \qquad E(x_+,x_-)
\longrightarrow E_0 \; .
\end{equation}
Relation (\ref{eq:J++}) reveals a linearly growing lightcone charge density,
\begin{equation}
J_+(x_+,x_-) \longrightarrow \frac12 J_0 x_+ \; ,
\end{equation}
just as our field theoretic computation produces for the case of a constant
electric field. The really interesting relation is (\ref{eq:J--}) which gives 
an {\it infinite} (and $x_-$ dependent) result for $J_-$,
\begin{equation}
J_-(x_+,x_-) \longrightarrow -\frac12 J_0 \left({1 + \beta \over 1 - \beta} x_+
+ x_-\right) \; .
\end{equation}

The physics of these results is quite simple. First note that any $e^+ e^-$
pair which is created with finite speed in the primed frame must be moving at 
the speed of light in the $-z$ direction after the infinite boost needed to
reach the unprimed frame. Recall that an on-shell particle has,
\begin{equation}
p^3 = {1 \over \sqrt{2}} \left(p_+ - {\widetilde{\omega}^2 \over 2 p_+}\right)
\; , 
\end{equation}
so $p^3 \rightarrow -\infty$ corresponds to $p_+ = 0^+$. This is why we only 
see particle production on the lightcone at $p_+ = 0$. Since electrons must 
accelerate in the $-z$ direction they immediately leave the manifold moving 
parallel to the $x_-$ axis. Positrons accelerate in the $+z$ direction so they 
stay on the manifold and, at late values of $x_+$, move parallel to the $x_+$ 
axis. This is why the lightcone charge density $J_+$ grows. The reason $J_-$ 
tends to be {\it infinite} is that both particles of each pair are created 
moving at the speed of light, so they contribute an infinite $p_-/p_+$. Of 
course they tend to cancel by virtue of their opposite charges. The reason 
$J_-$ is infinitely {\it negative} is that the electrons speed up while the 
positrons slow down. Finally, we can anticipate from the form of 
({\ref{eq:tprime}) that any nontrivial time dependence in the primed frame must
give rise to infinitely rapid evolution in $x_+$ on the unprimed frame.

Now consider the situation in the primed frame with back-reaction turned on.
What one sees at one loop is an approximately oscillatory electric field and
current \cite{Kluger}. Let us assume, for simplicity, that the behavior is 
exactly oscillatory and consistent with the Maxwell equation
$-A^{\prime\prime}(t') = J(t')$,
\begin{equation}
A(t') = {E_0 \over \omega} \sin(\omega t') \qquad , \qquad J(t') = \omega E_0
\sin(\omega t') \; .
\end{equation}
From relation (\ref{eq:vec}) one sees that the vector potential oscillates
infinitely fast with infinitely small amplitude,
\begin{equation}
\widehat{A}_-(x_+,x_-) \longrightarrow - \sqrt{1 - \beta} {E_0 \over \omega}
\sin\left({\omega x_+ \over \sqrt{1-\beta}}\right) \; .
\end{equation}
The electric field oscillates with the same amplitude as in the primed frame
but with infinite frequency,
\begin{equation}
E(x_+,x_-) \longrightarrow E_0 \cos\left({\omega x_+ \over \sqrt{1-\beta}}
\right) \; .
\end{equation}
From relation (\ref{eq:J++}) we see that $J_+$ goes to zero,
\begin{equation}
J_+(x_+,x_-) \longrightarrow \frac12 \sqrt{1 - \beta} \omega E_0 \sin\left(
{\omega x_+ \over \sqrt{1 - \beta}}\right) \; ,
\end{equation}
which means we do not need to consider vector potentials that depend upon $x_-$
in addition to $x_+$. Of course the source of the infinitely rapid oscillations
is the $J_-$ current which has infinite amplitude in addition to infinite
frequency,
\begin{equation}
J_-(x_+,x_-) \longrightarrow {-\omega E_0 \over \sqrt{1 - \beta}} \sin\left(
{\omega x_+ \over \sqrt{1 - \beta}}\right) \; .
\end{equation}
This all looks very much like what we found in the previous section.

\section{Discussion}

We have constructed a complete operator solution (\ref{eq:solution}) for free 
QED in the presence of an electric field that depends arbitrarily upon the
lightcone coordinate $x_+$. This class of backgrounds is general enough to 
include the actual evolution of the electric field as it changes due to
back-reaction from the current of electron-positron pairs which it induces. One
determines the actual electric field (to some order in the loop expansion) by 
computing the expectation value of $J_-$ (to this order), setting this equal to
$-A_-^{\prime\prime}(x_+)$, and solving the resulting equation. We did this to
one loop order in Section 6 and there is no essential obstacle to including 
higher loop effects. The vertices of QED do not even depend upon the
background, nor does the photon propagator. And with our operator solution we
have the essential elements of the electron propagator.

It might be useful to recapitulate the rather subtle way the equations of 
motion can be satisfied within our class of backgrounds. We started with the 
mode functions in a generic $A_-(x_+)$ gauge field. One consequence was 
expression (\ref{eq:firstform}) which states that the expectation value of 
$J_+$ grows with $e A_-(x_+)$. But then the Maxwell equation $\partial_- 
\partial_+ A_- = J_+$ implies that $A_-$ must depend on $x_-$, contradicting
our initial ansatz. The resolution of this apparent contradiction derives from
equation (\ref{eq:Maxppp}) for the renormalized expectation value of $J_-$. In
the limit that $\epsilon$ goes to zero the leading contribution to this source 
is negative infinite and independent of $x_-$. Hence so too is $\partial_+ E$.
In other words, having a finite, positive electric field causes the $x_+$
derivative of the electric field to become infinitely negative, which of course
drives the electric field to zero. At this point one of the assumptions of our
formalism breaks down, but it is easy to see, on physical grounds, that the
electric field must fall below zero and that the resulting negative electric 
field engenders a {\it positive} infinite $J_-$ current. This would lift it 
back up through zero, whereupon the (not necessarily periodic) cycle would
start again. Since the induced currents are infinite, the response time is 
zero. So the picture is of an electric field undergoing oscillations of finite 
amplitude with infinite frequency. Since our vector potential vanishes at $x_+ 
= 0$ we can recover it from the electric field by integration,
\begin{equation}
A_-(x_+) = - \int_0^{x_+} dy_+ E(y_+) \; .
\end{equation}
But this integral must vanish for an electric field undergoing oscillations of
finite amplitude with infinite frequency. Therefore our result 
(\ref{eq:firstform}) gives {\it zero} for the expectation value of $J_+$, and
there is no need for the solution to depend upon $x_-$.

One of the novel features of our solution is that the phenomenon of pair 
creation is a discrete event on the lightcone. Evolution is diagonal in the 
Fourier basis of $k_+$ and $\widetilde{k}$, however, it is the minimally
coupled, {\it kinetic} momentum $p_+ = k_+ - e A_-(x_+)$ which determines 
whether a particular Fourier component creates or annihilates particles at any 
given value of $x_+$. When $p_+$ passes from negative to positive that 
particular Fourier component experiences pair creation with probability $e^{-2
\pi\lambda(k_+,\widetilde{k})}$, where $\lambda$ is given by (\ref{eq:lambda}).
We exploited this at the end of Section 4 to give a simple and explicit 
derivation of the particle production rate per unit volume, in real time and
without resorting to {\it ad hoc} interpretations for formally meaningless
expressions. 

Why pair creation is so simple on the lightcone was explained in Section 6. 
Quantum field theory on surface of constant $x_+$ can be viewed as the infinite
boost limit of the conventional problem formulated on surfaces of constant $t'$
\cite{Kogut}. Pair production is not localized in time when the electric field
is homogeneous on surfaces of constant $t'$. Each of the various momentum modes
has a nonzero probability of appearing in {\it any} time interval. However, 
when subject to an infinite boost one sees that the newly created particles 
must appear, to the lightcone observer, to be moving with $p^3 \rightarrow -
\infty$. This corresponds to $p_+ \rightarrow 0^+$, which is why particles are 
created on the lightcone only when their kinetic momentum $p_+ = k_+ - e 
A_-(x_+)$ passes through zero.

Before a particular Fourier component undergoes pair production, the field at
$x_+$ is a mode function of modulus unity times the same Fourier component of
the field at $x_+ = 0$. After pair production the modulus of the mode function
drops by a factor of $e^{-\pi \lambda(k_+,\widetilde{k})}$. The missing 
amplitude is acquired by new operators which come in from $x_- = - \infty$. 
This may be one of the more interesting features of our solution for lightcone 
experts. It has long been known that specifying the fields on a surface of 
constant $x_+$ cannot completely determine their future evolution. This is 
obvious for massless fields in two spacetime dimensions. However, the problem 
has always been hidden at $k_+ = 0$ when either $m \neq 0$ or $D > 2$. It ever 
needed to be resolved if one only desired scattering amplitudes; these can be 
computed away from $k_+ = 0$ and then analytically continued. In our analysis 
the problem could not be avoided because more and more modes are pulled through
zero kinetic momentum $p_+ = k_+ - e A_-(x_+)$ as the long as the electric 
field remains positive.

Our original motivation for studying this problem was to see what it can teach
us about techniques for treating the related problem of quantum gravitational 
back-reaction on inflation. It is worth summarizing what we have learned in 
that context. First, there does not seem to be any generic problem with using 
expectation values to study back-reaction. The results we obtained by doing 
this in Section 6 have a transparently correct physical interpretation. We
should caution, however, that the current operator is a gauge invariant, unlike
the metric.

The second point of relevance is that back-reaction is an infrared effect. The
important physics is associated with the finite range of modes whose kinetic
momentum has passed through zero. We saw in Section 6 that the ultraviolet 
divergent contribution to the expectation value of $J_-$ comes from different
terms and has a different dependence upon the fields. Had we merely subtracted
these terms and replaced the bare charge and field everywhere with the 
renormalized ones we would have gotten the correct result. This {\it had} to 
work from the context of effective field theory, but it is comforting to see it
actually do so.

Finally, there is at least the possibility that one can follow the system into 
the regime where back-reaction is a strong effect. This can happen if the 1PI
diagrams past some finite order in the loop expansion make no large 
contribution to the effect. Then one will get the right result by simply
solving the effective field equations obtained by evaluating the expectation
value of the current operator to that finite order. Note especially that one 
does not have to simply do this and {\it hope} that it works. Once the solution
from the truncated effective field equations is obtained one can always check
to see whether the higher loop diagrams are in fact negligibly small in this
background. So the way is open to making a potentially self-consistent 
calculation.

\vskip 1cm
\centerline{\bf Acknowledgements}

We wish to acknowledge a stimulating conversation with D. Boyanovsky.
This work was partially supported by DOE contract DE-FG02-97ER\-41029, by
the Greek General Secretariat of Research and Technology grant 97 
E$\Lambda$-120, by EU grant HPRN-CT-2000-00122. The authors also express their
gratitude to the Institute for Fundamental Theory at the University of Florida
and to the Department of Physics at the University of Crete for hospitality
during mutual visits.


\begin{thebibliography}{99}

\bibitem{nct1} N. C. Tsamis and R. P. Woodard, Ann. Phys. {\bf 253}, 1 (1997).

\bibitem{abm} V. Mukhanov, L. R. W. Abramo and R. Brandenberger, Phys. Rev. 
Lett.  {\bf 78}, 1624 (1997); L. R. W. Abramo, R. H. Brandenberger and V. F.
Mukhanov, Phys. Rev. {\bf D56}, 3248 (1997).

\bibitem{atw} L. R. Abramo, R. P. Woodard and N. C. Tsamis, Fortshr. Phys. {\bf
47}, 389 (1999).

\bibitem{unruh} W. Unruh, ``Cosmological long wavelength perturbations,''
astro-ph/\-9802323.

\bibitem{Klein} O. Klein, Z. Phys. {\bf 53} 157 (1929).

\bibitem{Sauter} F. Sauter, Z. Phys. {\bf 69} 742 (1931) ; {\bf 73} 547 (1931).

\bibitem{Schwinger} J. Schwinger, Phys. Rev. {\bf 82}, 664 (1951).

\bibitem{Brezin} E. Brezin and C. Itzykson, Phys. Rev. {\bf D2}, 1191 (1970).

\bibitem{Casher} A. Casher, H. Neuberger and S. Nussinov, {\bf D20}, 179 (1979).

\bibitem{Bialynicki} I. Bialynicki-Birula, P. G\'ornicki, and J. Rafelski, 
Phys. Rev. {\bf D44}, 1825 (1991).

\bibitem{Kluger} Y. Kluger, J. M. Eisenberg, B. Svetitsky, F. Cooper, and E.
Mottola, Phys. Rev. {\bf D45}, 4659 (1992).

\bibitem{Best} C. Best and J. M. Eisenberg, Phys. Rev. {\bf D47}, 4639 (1993).

\bibitem{Gavrilov} S. P. Gavrilov and D. M. Gitman, Phys. Rev. {\bf D53}, 
7162 (1996).

\bibitem{Kluger2} Y. Kluger, J. M. Eisenberg, and E. Mottola, Phys. Rev. 
{\bf D58}, 125015 (1998).

\bibitem{Greiner} W. Greiner, B. M$\ddot {\rm u}$ller, and J. Rafelski,
{\it Quantum Electrodynamics of Strong Fields} (Springer-Verlag, Berlin, 1985)

\bibitem{Fradkin} E. S. Fradkin, D. M. Gitman, and S. M. Shvartsman,
{\it Quantum Electrodynamics with Unstable Vacuum} (Springer-Verlag, Berlin,
1991)

\bibitem{Indians1} K. Srinivasan and T. Padmanabhan, Phys. Rev. {\bf D16},
24007 (1999).

\bibitem{Indians2} K. Srinivasan and T. Padmanabhan, ``A novel approach to
particle production in a uniform electric field,'' gr-qc/9911022.

\bibitem{Wolkow} D. M. Wolkow, Z. Physik {\bf 94}, 250 (1935).

\bibitem{Kogut} J. B. Kogut and D. E. Soper, Phys. Rev. {\bf D1}, 2901 (1970).

\bibitem{Grad} I. S. Gradshteyn and I. M. Ryzhik, {\it Table Of Intergals 
Series And Products}, 4th Edition (Academic Press, New York, 1965), p. 937.

\end{thebibliography}
\end{document}